\documentclass[twocolumn,english,superscriptaddress,longbibliography, prx]{revtex4-1}
\usepackage[colorlinks=true,urlcolor=blue,citecolor=blue,linkcolor=blue]{hyperref} 
\usepackage[T1]{fontenc}
\usepackage[utf8]{inputenc}
\usepackage{amssymb}
\usepackage{graphicx}
\usepackage{amsmath,color}
\usepackage{mathrsfs}
\usepackage{float}
\usepackage{indentfirst}
\usepackage{txfonts}
\usepackage[normalem]{ulem}

\makeatletter

\def\journal #1, #2, #3, 1#4#5#6{{\sl #1~}{\bf #2}, #3 (1#4#5#6) }

\def\eqa{\begin{eqnarray}}
\def\eea{\end{eqnarray}}
\newcommand{\eq}{\begin{equation}}
\newcommand{\ee}{\end{equation}}

\newcommand{\Eq}[1]{Eq.~(\ref{#1})}

\newcommand{\Tr}{{\rm Tr}}

\newcommand{\circled}[1]{\raisebox{.5pt}{\textcircled{\raisebox{-.9pt} {#1}}}}

\makeatother

\usepackage{babel}

\begin{document}
\title{Differentiable Programming Tensor Networks}

\author{Hai-Jun Liao}
\affiliation{Institute of Physics, Chinese Academy of Sciences, Beijing 100190, China}
\affiliation{CAS Center for Excellence in Topological Quantum Computation, University of Chinese Academy of Sciences, Beijing 100190, China}
\author{Jin-Guo Liu}
\affiliation{Institute of Physics, Chinese Academy of Sciences, Beijing 100190, China}
\author{Lei Wang}
\email{wanglei@iphy.ac.cn}
\affiliation{Institute of Physics, Chinese Academy of Sciences, Beijing 100190, China}
\affiliation{CAS Center for Excellence in Topological Quantum Computation, University of Chinese Academy of Sciences, Beijing 100190, China}
\affiliation{Songshan Lake Materials Laboratory, Dongguan, Guangdong 523808, China}
\author{Tao Xiang}
\email{txiang@iphy.ac.cn}
\affiliation{Institute of Physics, Chinese Academy of Sciences, Beijing 100190, China}
\affiliation{University of Chinese Academy of Sciences, Beijing 100049, China}
\affiliation{Collaborative Innovation Center of Quantum Matter, Beijing 100190, China}

\begin{abstract}
Differentiable programming is a fresh programming paradigm which composes parameterized algorithmic components and trains them using automatic differentiation. The concept emerges from deep learning but is not only limited to training neural networks. We present theory and practice of programming tensor network algorithms in a fully differentiable way. By formulating the tensor network algorithm as a computation graph, one can compute higher order derivatives of the program accurately and efficiently using automatic differentiation. We present essential techniques to differentiate through the tensor networks contraction algorithms, including numerical stable differentiation for tensor decompositions and efficient backpropagation through fixed point iterations. As a demonstration, we compute the specific heat of the Ising model directly by taking the second order derivative of the free energy obtained in the tensor renormalization group calculation. Next, we perform gradient based variational optimization of infinite projected entangled pair states for quantum antiferromagnetic Heisenberg model and obtain start-of-the-art variational energy and magnetization with moderate efforts. Differentiable programming removes laborious human efforts in deriving and implementing analytical gradients for tensor network programs, which opens the door to more innovations in tensor network algorithms and applications. 
\end{abstract}
\maketitle


\section{Introduction} 
Tensor networks are prominent approaches for studying classical statistical physics and quantum many-body physics problems~\cite{Orus2014, Haegeman2016, Orus2018}. In recent years, its application has expanded rapidly to diverse regions include simulating and designing of quantum circuits~\cite{treewidth,Arad2010,Kim2017a,Huggins2019}, quantum error correction~\cite{Ferris2014a, Bravyi2014},  machine learning~\cite{Stoudenmire2016b, Stoudenmire2018g, Han2018, Cheng2019, Stokes}, language modeling~\cite{Gallego2017,Pestun2017a}, quantum field theory~\cite{Verstraete2010a,Haegeman2013,Hu2018,Tilloy2018} and holography duality~\cite{Swingle2012, Hayden2016}.

One of the central problems relevant to many research directions is the optimization of tensor networks in a general setting. Despite highly successful optimization schemes for one dimensional matrix product states~\cite{White1992, Vidal2006, Schollwock2011, Stoudenmire2011, Landau2015, Zauner-Stauber2018}, optimizing tensor networks in two or higher dimensions has been a challenging topic. The hardness is partly due to the high computational cost of tensor contractions, and partly due to the lack of an efficient optimization scheme in the high dimensional situation.

The difficulty is particularly pressing in optimizing tensor network states for infinite translational invariant quantum systems. In these cases, the same tensor affects the variational energy in multiple ways, which results in a highly nonlinear optimization problem. Optimization schemes based on approximate imaginary time projection~\cite{Jiang2008, Jordan2008, Corboz2010, Zhao2010, Phien2015} have been struggling to deal with nonlocal dependence in the objective function.
Reference~\cite{Corboz2016, Vanderstraeten2016} apply gradient based optimization and show it significantly improves the results. However, it is cumbersome and error prone to derive the gradients of tensor network states analytically, which involves multiple infinite series of tensors even for a simple physical Hamiltonian. This fact has limited broad adoption of the gradient based optimization of tensor network states to more complex systems. Alternative approaches, such as computing the gradient using numerical derivative has limited accuracy and efficiency, therefore only applies to cases with few variational parameters~\cite{Poilblanc2017, Chen2018k}. While deriving the gradient manually using the chain rule is only manageable for purposely designed simple tensor network structures~\cite{Wang2011}.

Differentiable programming provides an elegant and efficient solution to these problems by composing the whole tensor network program in a fully differentiable manner. In this paper, we present essential automatic differentiation techniques which compute (higher order) derivatives of tensor network programs efficiently to numeric precision. This progress opens the door to gradient-based (and even Hessian-based) optimization of tensor network states in a general setting. Moreover, computing (higher order) gradients of the output of a tensor network algorithm offer a straightforward approach to compute physical quantities such as the specific heat and magnetic susceptibilities. The differentiable programming approach is agnostic to the detailed lattice geometry, Hamiltonian, and tensor network contraction schemes. Therefore, the approach is general enough to support a wide class of tensor network applications.

We will focus on applications which involve two dimensional infinite tensor networks where the differentiable programming techniques offer significant benefits compared to the conventional approaches. We show that after solving major technical challenges such as numerical stable differentiation through singular value decomposition (SVD) and memory efficient implementation for fixed point iterations, one can obtain the state-of-the-art results in variational optimization of tensor network states. 

The organization of this paper is as follows. In section~\ref{sec:theory} we introduce automatic differentiation in the context of tensor network algorithms and formulate tensor network contractions as computation graphs. In section~\ref{sec:techniques} we present the key techniques for stable and scalable differentiable programming of tensor network algorithms. And section~\ref{sec:applications} we demonstrate the ability of the approach with applications to classical Ising model and quantum Heisenberg model on the infinite square lattice. Finally, we outlook for future research directions opened by this work in Sec.~\ref{sec:discussions}. Our code implementation is publicly available at~\cite{Github}. 

\section{General Theory} \label{sec:theory}
Automatic differentiation through a computation graph is a unified framework that covers training neural networks for machine learning, optimizing tensor networks for quantum physics, and many more. We first review the core idea of automatic differentiation and then explain its application to various tensor network contraction algorithms formulated in terms of computation graphs. 

\subsection{Automatic differentiation}\label{sec:AD}

Automatic differentiation mechanically computes derivatives of computation process expressed in terms of computer programs~\cite{Bartholomew-Biggs2000}. Unlike numerical differentiation, automatic differentiation computes the value of derivatives to the machine precision. 
The performance of automatic differentiation has a general theoretical guarantee, which does not exceed the algorithmic complexity of the original program~\cite{Baur1983,Griewank1989}.
Automatic differentiation is the computational engine of modern deep learning applications~\cite{BP1986,Baydin2018}.
Moreover, automatic differentiation also finds applications in quantum optimal control~\cite{Leung2017} and quantum chemistry calculations such as computing forces~\cite{Sorella2010} and optimizing basis parameters~\cite{Tamayo-Mendoza2018}.

\begin{figure}[t]
\includegraphics[width=\columnwidth]{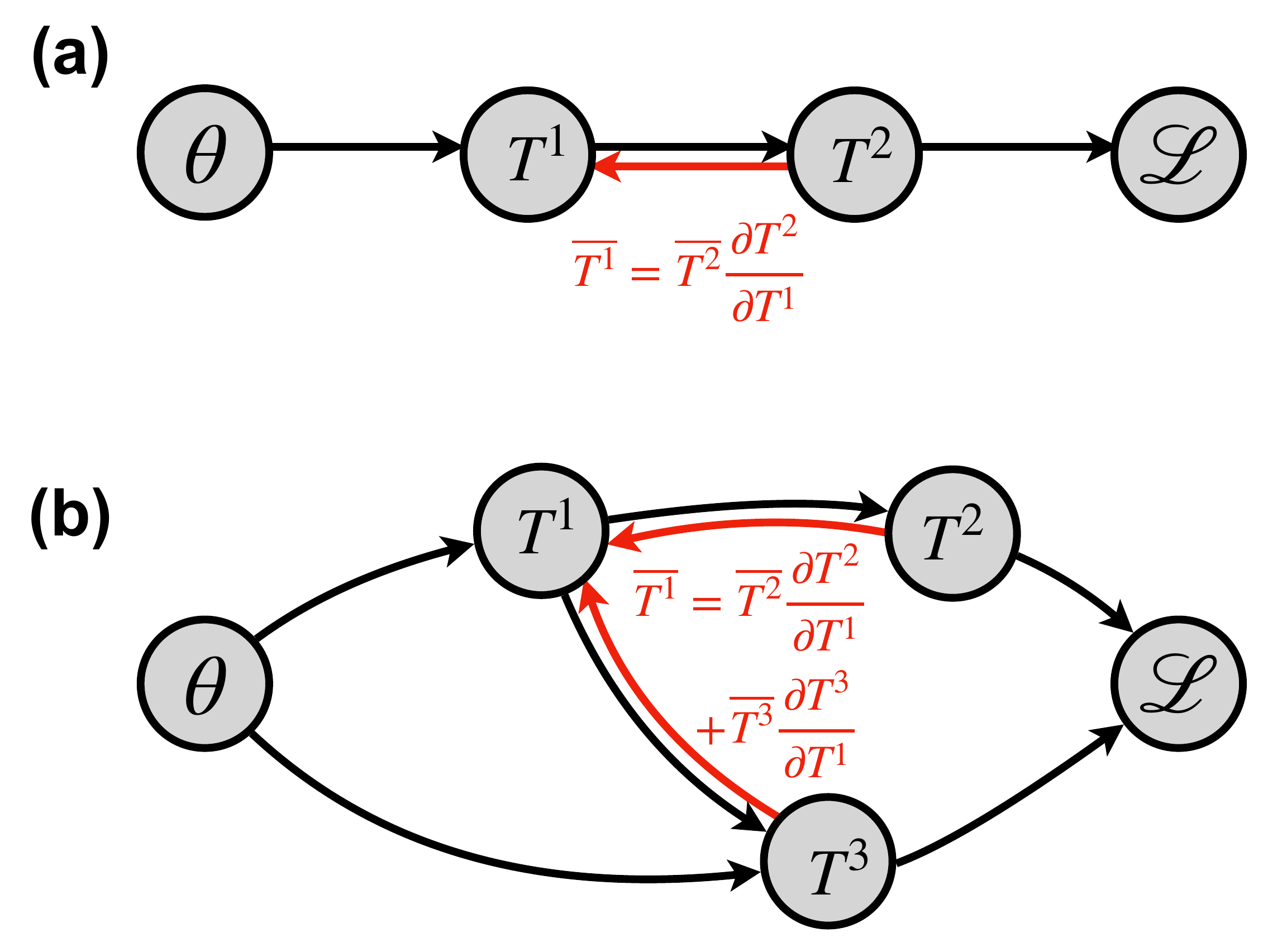}
\caption{Reverse mode automatic differentiation on computation graphs. Black arrows indicate the forward function evaluation from inputs to outputs. Red arrows indicate backward steps for adjoint backpropagation. (a) A chain graph. (b) A more general computation graph. In the backward pass, the adjoint of a given node is computed according to \Eq{eq:graphbp}.}
\label{fig:compgraph}
\end{figure}

Central to the automatic differentiation is the concept of the computation graph. A computation graph is a directed acyclic graph composed by elementary computation steps. The nodes of the graph represent data, which can be scalars, vectors, matrices, or tensors~\footnote{As noted in \cite{NIPS2018_7540} one will need tensor calculus as a precise language to present automatic differentiation, even the object being differentiated is a matrix.}. The graph connectivity indicates the dependence of the data flow in the computation process. The simplest computation graph is a chain shown in Fig~\ref{fig:compgraph}(a). Starting from, say vector valued, input parameters $\theta$ one computes a series of intermediate results until reaching the final output $\mathcal{L}$, which we assume to be a scalar. The so called forward evaluation simply traverses the chain graph in sequential order $\theta\rightarrow T^{1} \rightarrow \cdots \rightarrow T^{n} \rightarrow \mathcal{L}$. 

To compute the gradient of the objective function with respect to input parameters, one can exploit the chain rule
\begin{align}
\frac{\partial \mathcal{L}}{\partial \theta } = \frac{\partial \mathcal{L}}{\partial T^{n}}  \frac{\partial T^{n}}{\partial T^{n-1}} \cdots \frac{\partial T^{2}}{\partial T^{1}} \frac{\partial T^{1}}{\partial \theta}. 
\label{eq:chainrule}
\end{align}
Since we consider the case where the input dimension is larger than the output dimension, it is more efficient to evaluate the gradient in (\ref{eq:chainrule}) by multiplying terms from left to the right using a series of vector-Jacobian products. In terms of the computation graph shown in Fig.~\ref{fig:compgraph}(a), one traverses the graph backward and propagates the gradient signal from the output back to the input. Computing the derivative this way is called~\emph{reverse mode} automatic differentiation. This approach, commonly referred to as the backpropagation algorithm~\cite{BP1986}, is arguably the most successful method for training deep neural networks. 

It is instructive to introduce the \emph{adjoint variable} $\overline{T} = {\partial \mathcal{L}}/{\partial T}$ to denote the gradient of the final output $\mathcal{L}$ with respect to the variable $T$. One sees that the reverse mode automatic differentiation propagates the adjoint from $\overline{T^{n}} = \overline{\mathcal{L}}\,  \frac{\partial \mathcal{L}}{\partial T^{n}} $ with $\overline{\mathcal{L}}=1$ all the way back to $\overline{T^{i}} = \overline{T^{i+1}}  \frac{\partial T^{i+1}}{\partial T^{i}} $ with $i=n-1,\ldots, 1$, and finally computes $\overline{\theta} = \overline{T^1}\, \frac{\partial T^{1}}{\partial \theta} $. In each step, one propagates the adjoint backward via a local vector-Jacobian product.

The adjoint backpropagation picture generalizes well to more complex computation graphs. For example, the data node $T^1$ in Fig.~\ref{fig:compgraph}(b) affects the final output via two different downstream computation paths. 
In the backward pass, one needs to accumulate all contributions from its child nodes for its adjoint. In general, the backpropagation rule reads
\begin{equation}
\overline{T^{i}} = \sum_{j:\, \mathrm{child}\, \mathrm{of}\, i} \overline{T^{j}}  \, \frac{\partial T^{j}}{\partial T^{i}}. \label{eq:graphbp}
\end{equation} 

The reverse mode automatic differentiation algorithm can be understood as a message passing process on the computation graph. After a topological sort of the computation graph defined by the forward pass, one visits the graph backward from the output node with adjoint $\overline{\mathcal{L}}=1$. Each node collects information from its child nodes to compute its own adjoint, then passes this information to its parents. Thus, one can compute the gradient with respect to all parameters in one forward and one backward pass. Typically one caches necessary information in the forward pass for efficient evaluation of the vector-Jabobian product in the backward pass. 

The building blocks of a differentiable program are called \emph{primitives}. The primitives can be elementary operations such as addition, multiplication, and math functions~\cite{Griewank2008}. Each primitive has an associated backward function in addition to the ordinary forward function. The backward function backpropagates the adjoints according to the vector-Jacobian product \Eq{eq:graphbp}. Note that one does not need to explicitly instantiate or store the full Jacobian matrices. Moreover, one can group many elementary computation steps together as a primitive. For example, the linear algebra operations such as matrix multiplication can be regarded as a primitive. In this way, the forward pass of these customized primitive can be operated as a black box. Designing the differentiable program in such a modular way allows one to control the level of granularity of automatic differentiation. There are several advantages of crafting customized primitives for domain specific problems. First, this can reduce the computation and memory cost. Second, in some cases, it is numerically more stable by grouping several steps together. Third, one can wrap function calls to external libraries into primitives, without the need to track each individual computation step. 

Modern machine learning frameworks support automatic differentiation via various mechanisms. For example, \texttt{TensorFlow}~\cite{Abadi2016} explicitly constructs computation graphs using a customized language, \texttt{autograd}~\cite{ maclaurin2015autograd}, \texttt{PyTorch}~\cite{Paszke2017} and \texttt{Jax}~\cite{jaxpaper} track the program execution order at run time, and \texttt{Zygote}~\cite{innes2018don} performs source code transformation at the compiler level. These frameworks allow  one to differentiate through function calls, control flows, loops and many other computation instructions. Building on these infrastructures, \emph{differentiable programming} is emerging as a new programming paradigm that emphasizes assembling differentiable components and learning the whole program via end-to-end optimization~\cite{Baydin2018}. Letting machines deal with automatic differentiation mechanically has greatly reduced laborious human efforts and reallocated human attention to design more  sophisticated and creative deep learning algorithms. 

Finally, we note that it is also possible to evaluate \Eq{eq:chainrule} from right to left, which corresponds to the \emph{forward mode} automatic differentiation. The operation of the forward mode automatic differentiation is akin to the perturbation theory. One can compute the objective function and the gradient in a single forward pass without storing any intermediate results. However, the forward mode automatic differentiation is not favorable for computation graphs whose input dimension is much larger than the output dimension~\cite{Baydin2018}. Therefore, the majority of deep learning work employs the reverse mode automatic differentiation. For the same reason, we focus on the reverse mode automatic differentiation for tensor network programs. 

\subsection{Computation graphs of tensor network contractions}\label{sec:computationgraph}
A tensor network program maps input tensors to output tensors, which we assume to be a scalar. Depending on the contexts, the input tenors may represent classical partition function or quantum wavefunction, and the outputs can be various physical quantities of interest. 
It is conceptually straightforward to apply automatic differentiation to a tensor network program by expressing the computation, and in particular, the tensor network contractions as a computation graph.

As a pedagogic example, consider the partition function of infinite one dimensional Ising model $Z = \lim_{N\rightarrow \infty}\Tr(T^N)$ where  $T=\left(\begin{array}{cc} e^\beta & e^{-\beta} \\ e^{-\beta}  & e^\beta \end{array}\right)$ is a second rank tensor (matrix) representing the Boltzmann weight. One can numerically access the partition function in the thermodynamic limit by repeatedly squaring the matrix $T$ and tracing the final results.
The computation graph shows the simple structure as shown in Fig.~\ref{fig:compgraph}(a). Differentiating with respect to such computational process involves backpropagating the adjoint through matrix trace and multiplication operations, which is straightforward. 

At this point, it is also worth distinguishing between the exact and approximate tensor network contraction schemes. Tensor networks on tree like graph can be contracted exactly and efficiently. While other exact approaches, such as the one used for counting and simulating quantum computing~\cite{Kourtis2018, Villalonga2018}, in general exhibit exponentially scaling with the problem size. Nevertheless, it is straightforward to apply automatic differentiation to these exact algorithms since they mostly involve tensor reshape and contractions. 

We will focus on the less trivial cases of differentiating through \emph{approximate} tensor contractions, which typically involve truncated tensor factorization or variational approximations. They cover important tensor network  applications which show great advantages over other numerical methods~\cite{Orus2014, Haegeman2016, Orus2018}. In particular, we are interested in contracting infinite tensor network, where the fundamental data structure is the bulk tensor. The contraction schemes loosely fall into three categories, the ones based on coarse graining transformations~\cite{Levin2007, Xie2008, Gu2009, Xie2012, Xie2014, Evenbly2015, Yang2017c}, the ones based on the corner transfer matrix~\cite{Nishino1996, Orus2012, Corboz2014}, and the ones based on matrix product states~\cite{Vidal2006, Orus2008, Haegeman2016}. Since the last two contraction schemes are closely related~\cite{Haegeman2016}, in the following we will focus on automatic differentiation for the tensor renormalization group (Sec.~\ref{sec:trg}) and corner transfer matrix renormalization group approaches  (Sec.~\ref{sec:ctmrg}) respectively.

\begin{figure}[t]
\includegraphics[width=\columnwidth]{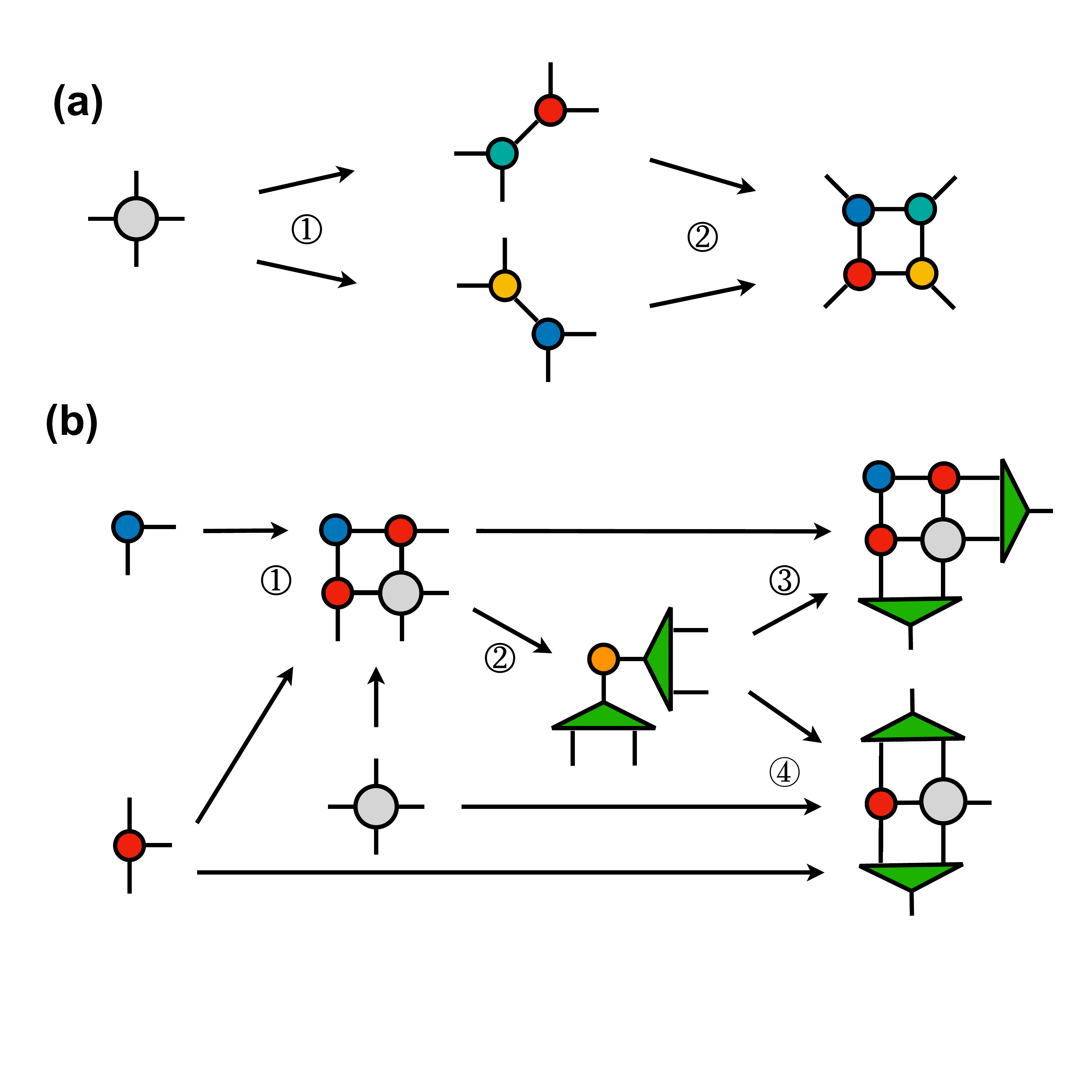}
\caption{(a) The iteration step of TRG. (b) The iteration step of CTMRG. Each tensor is a node in the computation graph. The primitive functions in the computation graphs are SVD and tensor contractions.}
\label{fig:rg}
\end{figure}

\subsubsection{Tensor renormalization group}\label{sec:trg}

Tensor renormalization group (TRG) contracts the tensor network by factorizing and blocking the bulk tensors iteratively~\cite{Levin2007}. Figure~\ref{fig:rg}(a) shows one step of the TRG iteration as the computation graph, which includes $\circled{1}$ Split the bulk tensor in two ways using SVD, where we have truncated the singular values and vectors to a prescribed bond dimension $\chi$. $\circled{2}$ Assemble the four 3-leg tensors generated in the last step into a 4-leg tensor. After this contraction, we obtain a new tensor for the next iteration. 

The TRG method grows the lattice size exponentially fast. So one quickly reaches the thermodynamic limit after a few tens of iterations. Note that for numerical stability one needs to rescale the tensor elements after each iteration. The computational cost of TRG method scales $\mathcal{O}(\chi^6)$ and the memory cost scales as  $\mathcal{O}(\chi^4)$. 
After unrolling the iterations, the computation graph of the TRG method is similar to the simple chain graph shown in Fig.~\ref{fig:compgraph}(a). Within each iteration step, the basic operations are tensor index permutation, truncated SVD and tensor contractions. Since each of these operations is differentiable, one can backpropagate through the TRG procedure to compute the derivative of a downstream objective function with respect to the input tensor. 

\subsubsection{Corner transfer matrix renormalization group}\label{sec:ctmrg}

The computation graph of the corner transfer matrix renormalization group (CTMRG)~\cite{Nishino1996} has a more interesting topology. The goal of CTMRG calculation is to obtain converged corner and edge tensors which represent the environment degrees of freedom of the bulk tensor. 

In cases where the bulk tensor has the full symmetry of the square lattice, the step of one CTMRG iteration is shown in Fig.~\ref{fig:rg}(b). $\circled{1}$  Contract the bulk tensor with the corner and edge tensors to form a 4-leg tensor. $\circled{2}$ Perform truncated SVD to the 4-leg tensor, keeping the singular dimensions up to the cut off $\chi$. Keep the truncated singular matrix as the isometric projector. $\circled{3}$ Apply the isometry to the 4-leg tensor from the first step to find a new corner tensor. $\circled{4}$ Apply the same isometry to find a new edge tensor for the next step. And iterate this procedure until convergence. One sees that the same bulk tensor with bond dimension $d$ appears in each step of the CTMRG iteration. Due to this reason, the converged environment tensors will depend on the bulk tensor in a complicated way. 

Unlike the TRG method~\cite{Levin2007}, the CTMRG approach grows the system size linearly. So one may need to iterate a bit more steps to reach convergences in CTMRG. On the other hand, the computational complexity $\mathcal{O}(d^3\chi^3)$ and memory cost $\mathcal{O}(d^2\chi^2)$ of CTMRG are smaller than the ones of TRG in terms of the cutoff bond dimension.

\section{Technical  Ingredients}\label{sec:techniques}

To compute gradients of a tensor network program using reverse mode automatic differentiation, one needs to trace the composition of the primitive functions and propagate the adjoint information backward on the computation graph. Thankfully, modern differentiable programming frameworks~\cite{maclaurin2015autograd, Abadi2016, Paszke2017, jaxpaper, innes2018don} have taken care of tracing and backpropagation for their basics data structure, differentiable tensors, automatically. 

What one needs to focus on is to identify suitable primitives of tensor network programs and define their vector-Jacobian products for backpropagation. The key components of tensor network algorithms are the matrix and tensor algebras. And there are established results on backward through these operations~\cite{Giles2008, Townsend, Seeger2017}. First of all, it is straightforward to wrap all~\texttt{BLAS} routines as primitives with customized backward functions. 
Next, although being less trivial, it is also possible to derive backward rules for many~\texttt{LAPACK} routines such as the eigensolver, SVD, and QR factorization~\cite{Giles2008}. By treating these linear algebra operations as primitives, one can compose a
differentiable program with efficient implementations of matrix libraries.

There are, however, a few practical obstacles to stable and scalable implementation of differentiable tensor network programs. First, the backward for the eigensolver and SVD may face numerical instability with degeneracy in the eigenvalues or singular values. Second, the reverse mode automatic differentiation may incur large memory consumption, which prevents one from reaching the same bond dimension of an ordinary tensor network program. We present solutions to these problems in below. 

\subsection{Stable backward through linear algebra operations} \label{sec:la}

We present several key results on matrix derivatives involving linear algebra operations that are relevant to tensor network algorithms. Recall the modular nature of reverse mode automatic differentiation, one just needs to specify the local backward function to integrate these components into a differentiable program. We will comment on their connections to physics literature and pay special attention to stable numerical implementations~\cite{Github}. For more information, one can refer to~\cite{Giles2008, Townsend, Seeger2017}.  

\subsubsection{Symmetric eigensolver} \label{sec:symeig}
The forward pass reads $A = U D U^T$, where the diagonal matrix $D$ is a diagonal matrix of eigenvalues $d_i$ and each column of the orthogonal matrix $U$ is a corresponding eigenvector. In the computation graph the node $A$ has two child nodes $U$ and $D$.

In the backward pass, given the adjoint $\overline{U}$ and $\overline{D}$, we have~\cite{Giles2008, Seeger2017}
\begin{equation}
\overline{A} = U \left[\overline{D} +  F \odot  (U^T\overline{U} - \overline{U}^T U)/2 \right] U^T, 
\label{eq:symeig_backward}
\end{equation}
where $F_{ij}=(d_j - d_i)^{-1}$ if $i\neq j$ and zero otherwise. The symbol $\odot$ denotes an element-wise Hadamard product. One can readily check that the gradient is also a symmetric matrix. 

Equation~(\ref{eq:symeig_backward}) can be regarded as "reverse" perturbation theory. When the downstream calculation does not depend on the eigenstate, i.e. $\overline{U}=0$, the backward equation is related to the celebrated Hellmann-Feynman theorem ~\cite{Feynman1939} which connects the perturbation to the Hamiltonian and its eigenvalues. Ref.~\cite{Fujita2018} applied this special case of \Eq{eq:symeig_backward} for inverse Hamiltonian design based on energy spectra.

The appearance of the eigenvalue difference in the denominator of $F$ is a reminder of the first order nondegenerate perturbation theory. Reference~\cite{Tamayo-Mendoza2018} concerned about the stability of the backward through the eigensolver, thus turned to less efficient forward mode automatic differentiation for variational optimization of Hartree-Fock basis. 
Actually in many physical problems the final object function depends on part of the eigenvalues and eigenstates in a gauge independent way, e.g., a quadratic form of occupied eigenstates. In these cases, only the eigenvalue difference between the occupied and unoccupied states will appear in the denominator of $F$, which is a familiar patten in the linear response theory~\cite{martin2004electronic}. Therefore, degenerated eigenvalues would not necessarily cause problem for these physical applications~\footnote{The cases where one has exact degeneracy between occupied and unoccupied eigenstates is a singular case. One should include or exclude all degenerated states in the occupation.}. 
In practice, we found that by using a Lorentzian broadening with $1/x \rightarrow x/(x^2+\varepsilon)$ with $\varepsilon=10^{-12}$, one can stabilize the calculation at the cost of introducing a small error in the gradient, see also \cite{Seeger2017}. 

\subsubsection{Singular value decomposition}\label{sec:svd}
A ubiquitous operation in tensor network algorithms is the matrix SVD, which is used for canonicalization and factorization of tensor networks~\cite{Schollwock2011, Orus2014, Haegeman2016}. The forward pass reads  $A= U D V^T$, where $A$ is of the size $(m, n)$, and $U, V^T$ has the size $(m, k)$ and $(k, n)$ respectively, and $k=\min(m, n)$. $D$ is a diagonal matrix contains singular values $d_i$. In the reverse mode automatic differentiation, given the adjoints $\overline{U}, \overline{D}$ and $\overline{V}$, one can obtain~\cite{Townsend}
\begin{align}
\overline{A} & = \frac{1}{2} U\left[ F_{+}\odot \left(U^T \overline{U} -\overline{U}^T U \right) + F_{-} \odot \left(V^T \overline{V} -\overline{V}^T V \right)  \right ] V^T \nonumber \\
 & + U  \overline{D} V^T + (I-UU^T)\overline{U}D^{-1} V^T + UD^{-1}\overline{V}^T(I-VV^T), \label{eq:SVD_backward}
\end{align}
where $[F_\pm]_{ij}= \frac{1}{d_j - d_i} \pm \frac{1}{d_j + d_i} $ for $i\neq j$ and zero otherwise. To prevent the numerical issue in case of degenerate singular values, we use the same Lorentzian broadening as Sec.~\ref{sec:symeig} for the first term, which works well in our experience. In practice, for variational tensor network calculation starting from random tensors, the chance of having exact degenerate eigenvalues is small. And even if this happens, applying the rounding is a reasonable solution. While for the cases of degeneracy due to intrinsic reasons~\cite{Gu2009, Pollmann2010, Xie2014, Efrati2014}, one will still obtain the correct gradient as long as the end-to-end gradient is well defined. Lastly, inverting the vanishing singular values in \Eq{eq:SVD_backward} is not a concern since the corresponding space is usually truncated.


\subsubsection{QR factorization}
QR factorization is often used for canonicalization of tensor networks~\cite{Schollwock2011, Orus2014,  Haegeman2016}.  In the forward pass, one factorizes $A = QR$, where $Q^TQ =I$ and $R$ is an upper triangular matrix~\footnote{One can also require the diagonal of the $R$ matrix to be positive 
to fix the redundant gauge degrees of freedom. This can be easily implemented on top of the existing QR libraries with backward support.}. 
Depending on the dimensions $(m, n)$ of the matrix $A$ there are two cases for the backward function. 

For input shape of $A$ matrix $m\geq n$, $R$ is a $n\times n$ matrix. The backward pass reads~\cite{Seeger2017} 
\begin{equation}
\overline{A} = \left[ \overline{Q} + Q\, \texttt{copyltu}(M)\right]R^{-T},\label{eq:qrbp1}
\end{equation}
where $M = R\overline{R}^T- \overline{Q}^TQ$ and the \texttt{copyltu} function generates a symmetric matrix by copying the lower triangle of the input matrix to its upper triangle, $[\texttt{copyltu}(M)]_{ij}=M_{\max(i,j), \min(i,j)}$. The multiplication to $R^{-T}$ can be dealt with by solving a linear system with a triangular coefficient matrix. 

For the case of $m<n$, $Q$ is of the size $(m, m)$, and $R$ is a $m\times n$ matrix.  We denote $A = (X, Y)$ and $R=(U, V)$, where $X$ and $U$ are full rank square matrices of size $(m, m)$. This decomposition can be separated into two steps, first $X=QU$ uniquely determines $Q$ and $U$, then we calculate $V=Q^TY$. Applying the chain rule, the backward rule gives
\begin{equation}
\overline{A} =\left( \left[ (\overline{Q}+\overline V Y^T) + Q\, \texttt{copyltu}( M)\right] U^{-T}, Q\overline{V}\right), \label{eq:qrbp2}
\end{equation}
where $M = U\overline{U}^T-(\overline{Q}+\overline{V}Y^T)^TQ$. 

\subsection{Memory efficient reverse mode automatic differentiation with checkpointing function} \label{sec:checkpoint}
A straightforward implementation of the reverse mode automatic differentiation 
for tensor networks has a large memory overhead. This is because one needs to store intermediate results in the forward pass for evaluating the vector-Jacobian products in the backward pass. The number of stored variables is related to the level of granularity in the implementation of the automatic differentiation. In any case, the memory consumption of reverse mode automatic differentiation will be proportional to the depth of the computation graph. This is particularly worrying for tensor networks with large bond dimensions and a large number of renormalization iterations. 

The solution to the memory issue of reverse mode automatic differentiation is a well known technique called checkpointing~\cite{Griewank2008, chen2016training}. The idea is to trade the computational time with memory usage. Taking the chain computation graph in \Eq{eq:chainrule} as an example, one can store the tensor every a few steps in the forward process. And in the backward pass, one recomputes intermediate tensors whenever needed by running a small segment of the computation graph forwardly. In this way, one can greatly reduce the memory usage with no more than twice of the computational effort. 

In a deeper understanding, checkpointing amounts to define customized primitives which encapsulates a large part of the computation graph. These primitives have their own special backward rules which locally runs the forward pass again and then backpropagates the adjoint. Therefore, in the forward pass one does not need to cache internal states of these checkpointing primitives.

The checkpointing is a general strategy that is applied to the computation graph of any topological structure. When applied to tensor network algorithms, it is natural to regard the renormalization steps shown in Fig.~\ref{fig:rg} as checkpointing primitives. In this way, one avoids storing some large intermediate tensors in the forward pass. 

\subsection{Backward through fixed point iteration \label{sec:fixedpoint}} 

Fixed point iteration is a recurring pattern in tensor network algorithms. For example, one iterates the function $T^{i+1} = f(T^{i}, \theta)$ until reaching a converged tensor $T^\ast$ and uses it for downstream calculations. To compute the gradient with respect to the parameter $\theta$, one can certainly unroll the iteration to a deep computation graph and directly apply the reverse mode automatic differentiation. However, this approach has the drawback of consuming large memory if it takes long iterations to find the fixed point. 

One can solve this problem by using the implicit function theorem on the fixed point equation  ~\cite{Christianson1994}. Taking the derivative on both sides of $T^{\ast} = f(T^{\ast}, \theta)$, we have
\begin{align}
\overline{\theta} =  \overline{T^\ast}\, \frac{\partial T^\ast}{\partial \theta} & = \overline{T^\ast}\left[I- \frac{\partial f(T^\ast, \theta)}{\partial T^\ast}\right]^{-1} \frac{\partial f(T^\ast, \theta)}{\partial \theta} \nonumber \\
& = \sum_{n=0}^\infty \overline{T^\ast} \left[\frac{\partial f(T^\ast, \theta)}{\partial T^\ast}\right]^n \frac{\partial f(T^\ast, \theta)}{\partial \theta}. 
\label{eq:fixedpoint}
\end{align}
The second line expands the matrix inversion in the square bracket as a geometric series. Therefore, to backpropagate through a fixed point iteration, the basic operation is just the vector-Jacobian products involving the single step iteration function. And in the backward function one performs iteration to  accumulate the adjoint $\overline{\theta}$ until reaching its convergence. The geometric series show the same convergence rate to the fixed point as the forward iteration~\cite{Christianson1994}. 

Many of the tensor network contraction schemes, including the CTMRG method reviewed in Sec.~\ref{sec:ctmrg}, fall into the framework of fixed point iterations. Thus, one can use \Eq{eq:fixedpoint} for backward through CTMRG calculation, where the iteration is over the RG step shown in Fig.~\ref{fig:rg}(b). We note  that the analytical gradient of infinite tensor network contraction derived in Refs.~\cite{Corboz2016, Vanderstraeten2016} contains similar pattern, which is a summation of geometric series. 

Similar to the checkpoint technique of Sec.~\ref{sec:checkpoint}, \Eq{eq:fixedpoint} also reduces the memory usage in the reverse mode automatic differentiation since one does not need to store a long chain of intermediate results in the forward iteration.
Moreover, since the downstream objective function is independent of how the fixed point tensor is obtained, one can also exploit an accelerated iteration scheme~\cite{Haegeman2018a} in the forward process~\footnote{A related example is backward through an iterative solver for the dominant eigenstates $x^\ast$ of a matrix $A$. In the forward process, one may use the Krylov space method such as Lanczos or Arnoldi iterations. While for the backward pass one can leverage the fact that the dominant eigenvector is the fixed point of the power iteration $ f(x, A) = Ax/||A x||$. Therefore, one can use \Eq{eq:fixedpoint} to propagate $\overline{x^\ast}$ back to $\overline{A}$. Note that the forward and the backward iteration functions are decoupled, thus the backward function does not need to be the same as the forward pass, as long as it ensures $x^\ast$ is a fixed point.}. There is, however, a caveat when applying \Eq{eq:fixedpoint} to differentiating tensor network RG algorithms. One may need to pay special attention to the redundant global gauge in the RG iterations to ensure the fixed point equation indeed holds. 

\subsection{Higher order derivatives}\label{sec:higherorder}
Since the gradient can also be expressed as a computation graph, one can compute the seconder order derivatives by applying automatic differentiation to the graph again. In this way, one can in principle compute arbitrary higher order derivatives of a program using automatic differentiation~\cite{Griewank2008}. Deep learning frameworks~\cite{maclaurin2015autograd, Abadi2016, Paszke2017, jaxpaper, innes2018don} have out-of-the-box support for computing higher order derivatives. 

The ability to compute higher derivatives supports Hessian based optimization of tensor network states, such as the Newton method~\cite{Nocedal2006}, for tensor network states. However, computing and inverting the full Hessian matrix explicitly could be prohibitively expensive and unnecessary. One can efficiently compute the Hessian-vector product via $\sum_j  \frac{\partial ^2\mathcal{L}}{\partial \theta_i \partial \theta_j}   x_j = \frac{\partial }{\partial \theta_i} \left(\sum_j  \frac{\partial \mathcal{L}}{\partial \theta_j} x_j\right )$ without constructing the Hessian matrix explicitly~\cite{pearlmutter1994fast}. This is sufficient for iterative linear equation solvers used for the Newton method. 

\section{Applications}\label{sec:applications}
We present two applications to demonstrate the versatility of differentiable programming tensor network approach for statistical physics and quantum many-body problems. Our public available code implementation~\cite{Github} employs \texttt{PyTorch}~\cite{pytorch} with a customized linear algebra automatic differentiation library for improved numerical stability, see discussions in Sec.~\ref{sec:la}. While we note that one is readily to reproduce the results with other modern deep learning frameworks such as~\texttt{autograd}~\cite{autograd}, \texttt{TensorFlow}~\cite{tensorflow}, \texttt{Jax}~\cite{Jax}, and \texttt{Zygote}~\cite{Zygote} frameworks. 

\subsection{Higher order derivative of the free energy}\label{sec:Ising}

\begin{figure}[t]
\includegraphics[width=\columnwidth]{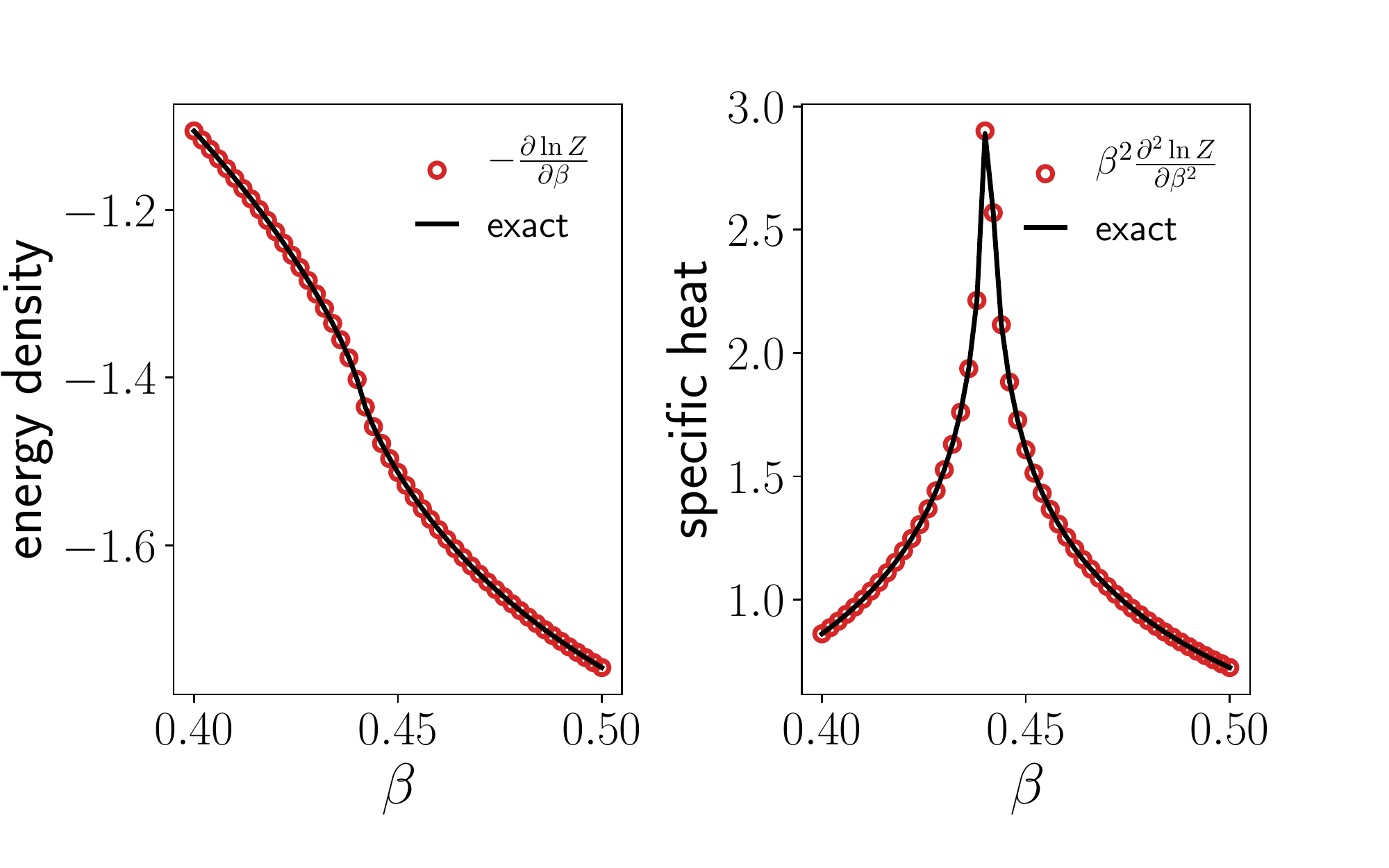}
\caption{Energy density and specific heat of the 2D Ising model. 
They are computed by taking the first and second order derivative of the free energy obtained after $30$  TRG iteration steps with a cutoff bond dimension $\chi=30$. Solid lines are exact solutions~ \cite{Onsager1944}.}
\label{fig:Ising}
\end{figure}

Consider an Ising model on the square lattice with inverse temperature $\beta$, its partition function can be expressed as a two dimensional tensor network with bond dimension $D=2$
\begin{align}
    Z = \raisebox{-8ex}{\includegraphics[scale=0.3, trim={1cm 1cm 0 0}, clip]{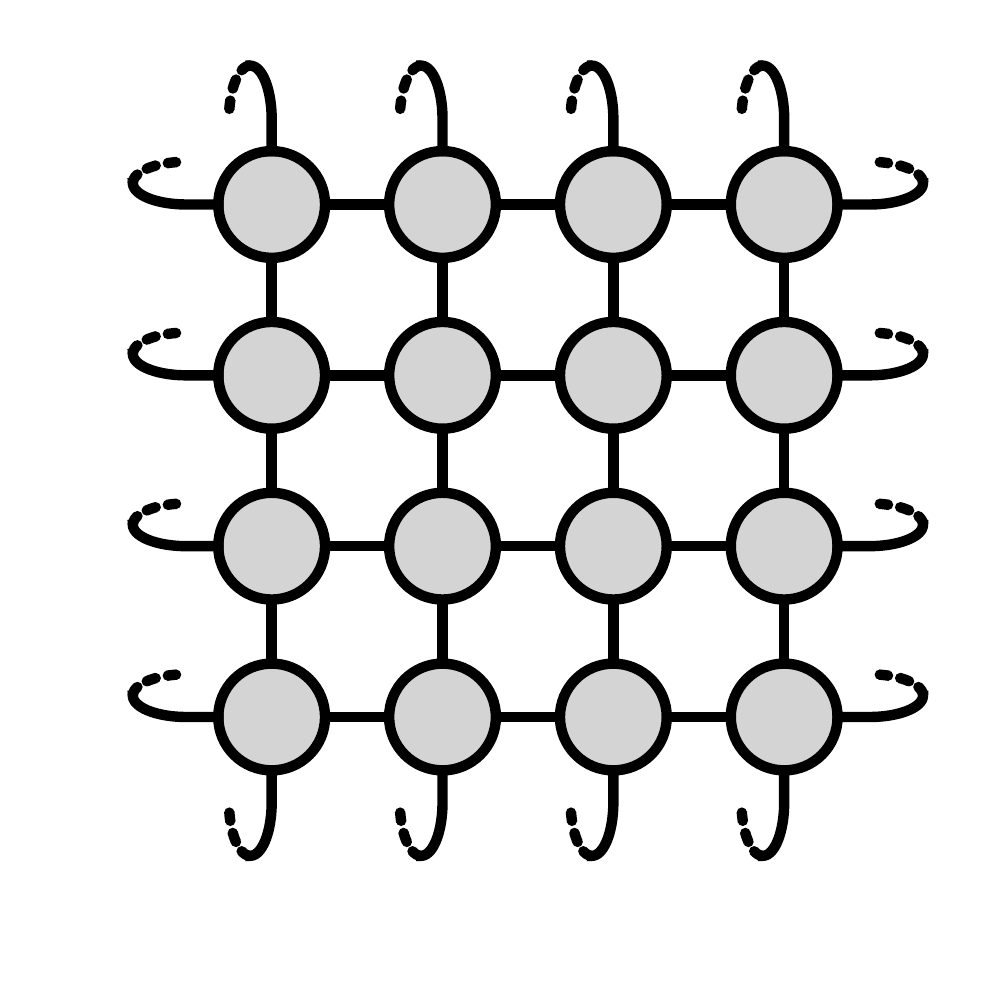}}. 
    \label{eq:IsingZ}
\end{align}
The bulk tensor is~\cite{q-clock}
\begin{align}
T_{uldr} =\raisebox{-3ex}{\includegraphics[scale=0.3, trim={1.7cm 1cm 0 0}, clip]{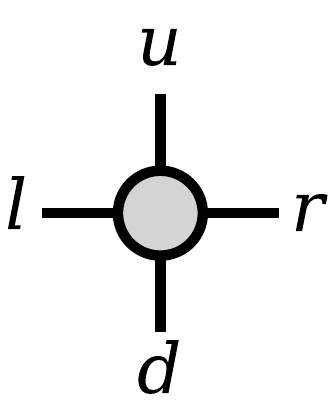}} = \frac{\sqrt{\lambda_u\lambda_l\lambda_d\lambda_r}}{2}\,\,\delta_{\mathrm{mod}(u+l-d-r,2)}, 
\end{align}
where $\lambda_u = e^\beta + (-1)^u e^{-\beta}$. We contract the infinite tensor network using the TRG approach discussed in Sec.~\ref{sec:trg}. We use a cut off bond dimension $\chi=30$ and iterate for $30$ TRG steps. Finally, we obtain the partition function \Eq{eq:IsingZ} and the free energy by tracing out the bulk tensor.  


Next, we compute the physical observables such as energy density and specific heat by directly taking derivatives of the free energy using automatic differentiation, as shown in Fig.~\ref{fig:Ising}. One notices that the energy density shows a kink and the specific heat exhibits a peak around the critical temperature $\beta_c=\ln(1+\sqrt{2})/2\approx 0.44068679$. Unlike numerical differentiation, these results are free from the finite difference error~\cite{Xie2012, Li2012}. Accurate computation of higher order derivatives of the tensor network algorithm will be useful to investigate thermal and quantum phase transitions. We note that it is in principle possible to obtain the specific heat by directly computing the energy variance~\cite{Vanderstraeten2015, Vanderstraeten2016}, which, however, involves cumbersome summation of geometric series expressed in term of tensor networks. 

There are alternative ways to compute the specific heat with automatic differentiation. For example, one can directly compute the energy via using the impurity tensor and then take the first order derivative to obtain the specific heat. Or, one can also use forward mode automatic differentiation since there is only one input parameter $\beta$ to be differentiated. We have purposely chosen the present approach to show off the power of differentiable programming with the reverse mode automatic differentiation technique. Backpropagating through the whole TRG procedure, and in particular the SVD, allows one to compute physical observables using higher order derivatives. It is remarkable that this works at all given many of the degenerate singular values due to the $Z_2$ symmetry of the Ising model~\cite{Tamayo-Mendoza2018}. To obtain correct physical results, it is crucial to implement the SVD backward function in a numerical stable way as explained in Sec.~\ref{sec:svd}. 

\subsection{Gradient based optimization of iPEPS}

We consider a variational study of the square lattice antiferromagnetic Heisenberg model with the Hamiltonian
\begin{equation}
H = \sum_{\langle i ,j \rangle} S^x_i S^x_j +  S^y_i S^y_j + S^z_i S^z_j.
\end{equation}
We consider an infinite projected entangled pair state (iPEPS) as the variational ansatz. The variational parameters are the elements in the iPEPS 
\begin{align}
    A^{s}_{uldr}=\raisebox{-3ex}{\includegraphics[scale=0.3, trim={1cm 1cm 0 0}, clip]{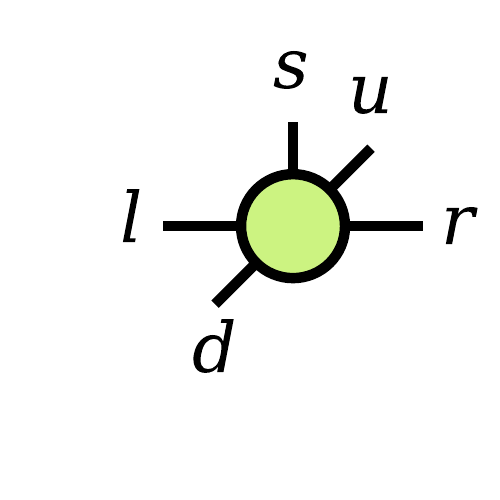}},  \label{eq:singlelayer}
\end{align}
where $s$ denotes the physical indices, and the remaining indices $u,l,d,r$ are for virtual degrees of freedom of the bond dimension $D$. We initialize the tensor elements with random Gaussian variables. The overlap of the iPEPS forms a tensor network, where the bulk tensor is the double layer tensor with bond dimension $d=D^2$
\begin{align}
T_{uldr} = \raisebox{-3ex}{\includegraphics[scale=0.3, trim={1.7cm 1cm 0 0}, clip]{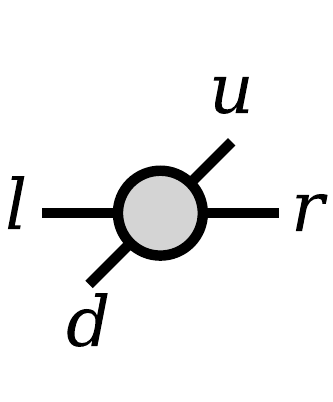}} = \raisebox{-3ex}{\includegraphics[scale=0.3, trim={2cm 1cm 0 0}, clip]{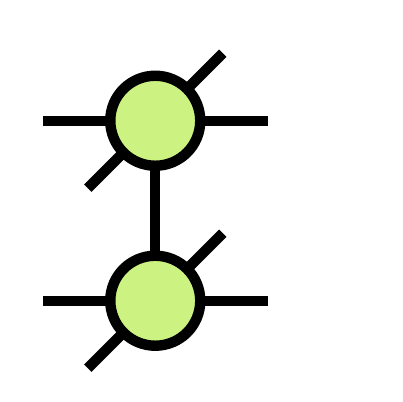}}.  
\end{align}
To contract the infinite tensor network formed by this bulk tensor we use the CTMRG method reviewed in Sec.~\ref{sec:ctmrg}. We initialize the corner and edge tensors by partially tracing out legs from the bulk tensor, then perform the CTMRG iteration until we reach convergence in the corner and edge tensors. 
After contraction, we can evaluate the expected energy ${\langle \psi | H | \psi \rangle}/{\langle \psi | \psi \rangle}$. Due to the translational invariance of the problem, it is sufficient to consider the expected energy on a bond
\begin{align}
\mathcal{L} = {\raisebox{-3.5ex}{\includegraphics[scale=0.25, trim={1.8cm 3cm 0 0}, clip]{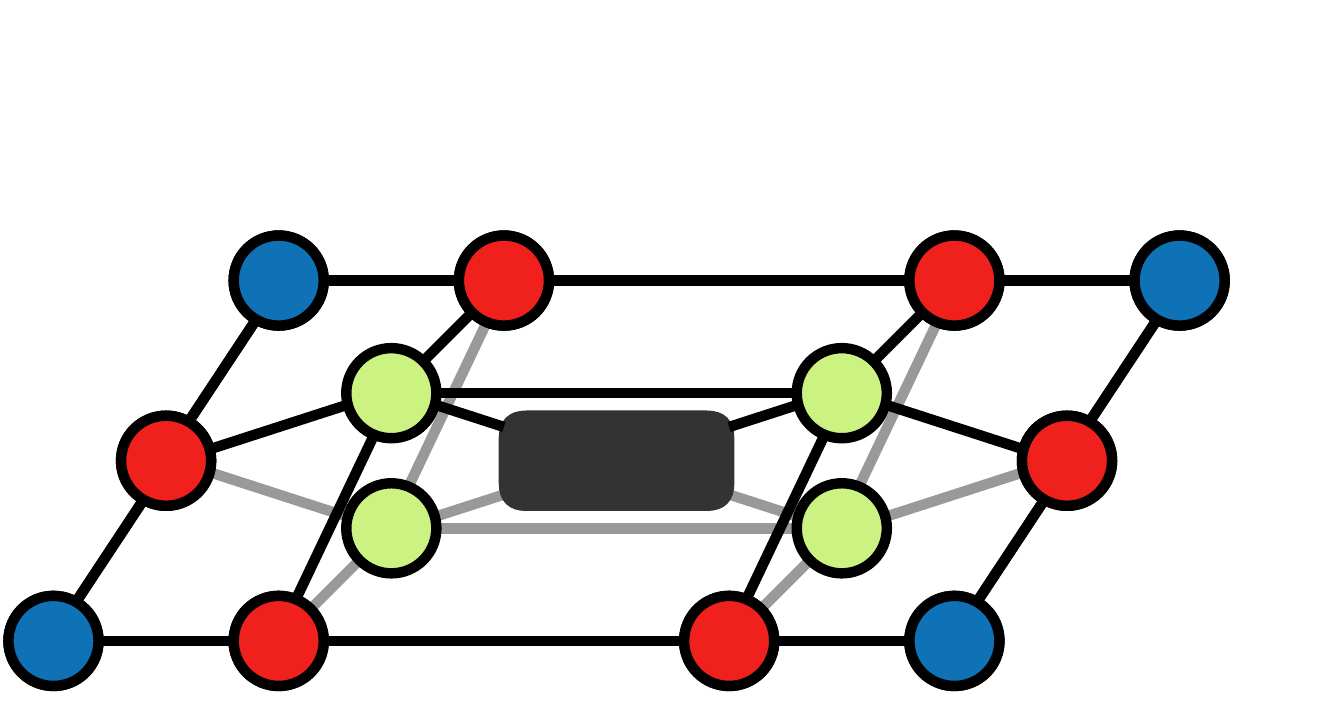}}} \Bigg/
{\raisebox{-3.5ex}{\includegraphics[scale=0.25, trim={1.8cm 3cm 0 0}, clip]{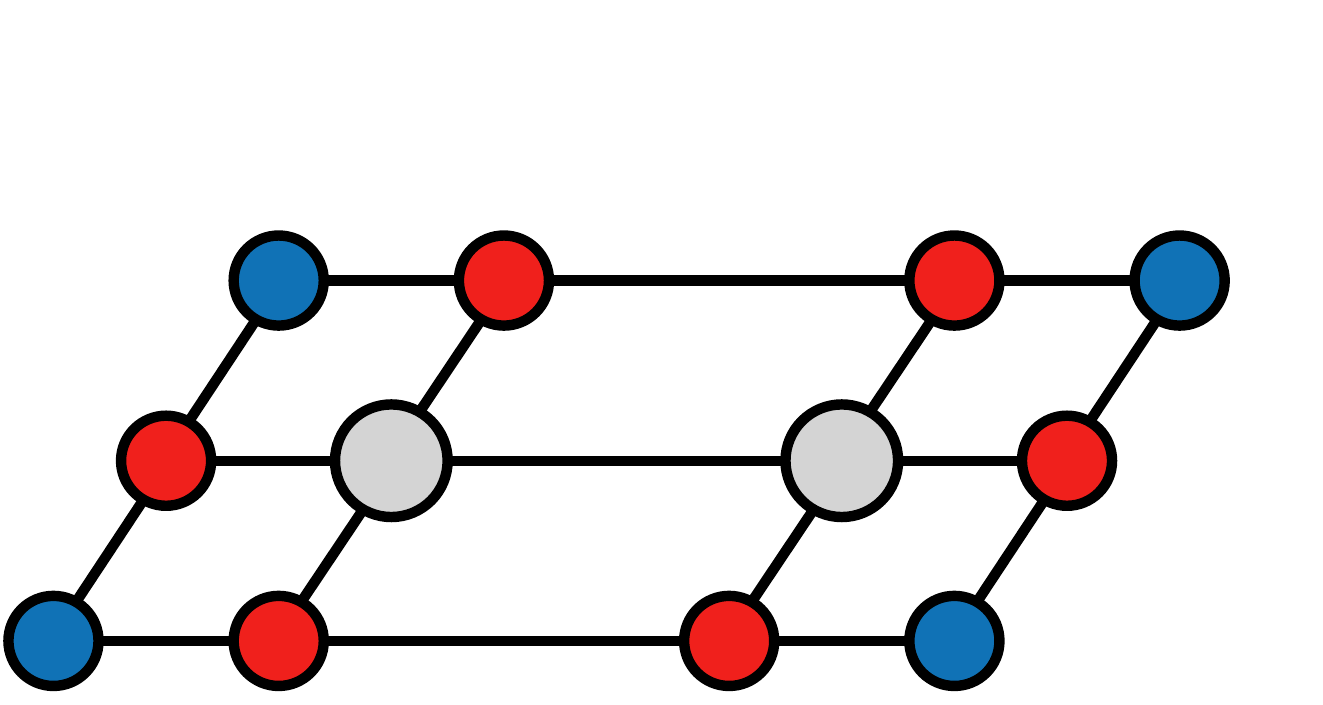}}} \label{eq:Hbond},  
\end{align}
where the black rectangle in \Eq{eq:Hbond} is the Hamiltonian operator acting on a bond. We have performed a basis rotation to the Hamiltonian so that the ground state will have a single site unit cell. We use cutoff bond dimension $\chi=30,50,80,100,144,160$ for $D=2,3,\ldots,7$ respectively. Since the expected energy decreases with the cutoff dimension~\cite{Poilblanc2017, Xie2017}, the approximated CTMRG contraction gives variational upper bound to the ground state energy. The expected energy \Eq{eq:Hbond} has both explicit and implicit dependence on the variational parameters in \Eq{eq:singlelayer} via the corner and edge tensors.  

\begin{figure}[t]
\includegraphics[width=0.5\textwidth,clip=true,trim={1cm 0 0 0}]{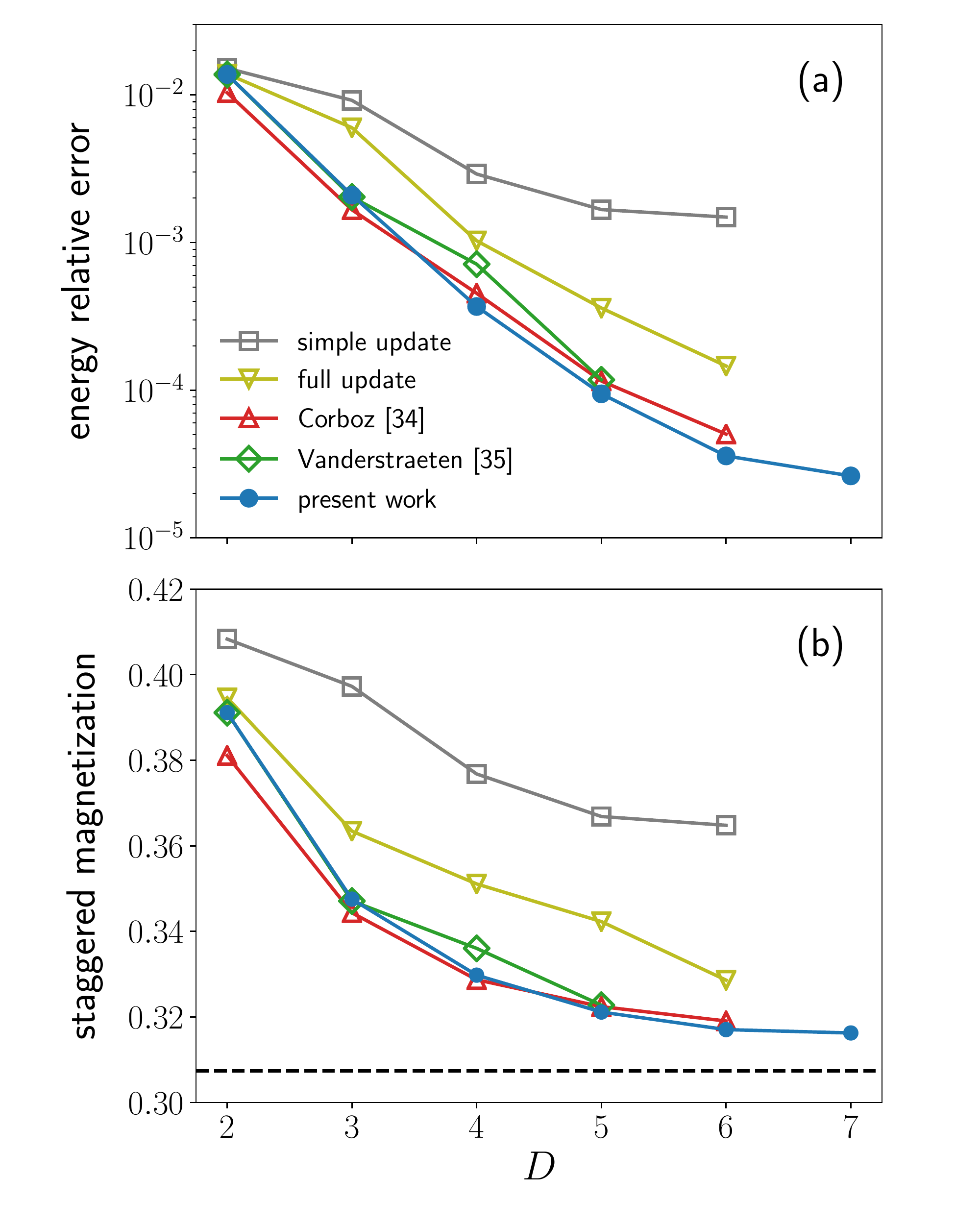}
\caption{(a) The relative error in the energy of 2D $S=1/2$ antiferromagnetic Heisenberg model
compared to previous variational results~\cite{Corboz2016,Vanderstraeten2016}. The accuracy is measured relative to the extrapolated quantum Monte Carlo (QMC) result~\cite{Sandvik2010a}. (b) A comparison of the staggered magnetization, where the dashed line is the extrapolated QMC result~\cite{Sandvik2010a}. The simple and full update reference data are also from Ref.~\cite{Corboz2016}. }
\label{fig:EM_Heis}
\end{figure}

We compute the gradient of \Eq{eq:Hbond} with respect to the single layer tensor \Eq{eq:singlelayer} using automatic differentiation, which automatically resolves the intricate structure in the computation graph Fig.~\ref{fig:rg}(b). The gradient computation takes time comparable to the forward evaluation of the expected energy. Then, we optimize the iPEPS using quasi-Newton L-BFGS algorithm~\cite{Nocedal2006} with the automatically computed gradient. One quickly reaches an optimum after a few hundred function and gradient evaluations. Figure~\ref{fig:EM_Heis}(a) shows the relative error in energy compared to extrapolated quantum Monte Carlo (QMC) results~\cite{Sandvik2010a} for various bond dimensions. The accuracy of the ground state energy is comparable to the state-of-the-art results~\cite{Corboz2016, Vanderstraeten2016}, which were shown to be more accurate than imaginary time projection based simple and full update algorithms~\cite{Jiang2008, Jordan2008, Corboz2010, Zhao2010, Phien2015} \footnote{Although it was expected that the full update method will in principle reach the same accuracy as the variational method within the error of Trotter-Suzuki decomposition, it would require optimizing the iPEPS tensors in a globally optimal way.}.
Note that both \cite{Vanderstraeten2016} and our ansatz contain only half of the variational parameters of the one in~\cite{Corboz2016}, so the energy results are slightly higher than Ref.~\cite{Corboz2016} at $D=2,3$. However, for larger bond dimensions $D=4,5,6,7$, our calculation reaches the lowest variational energy for the infinite square lattice Heisenberg model. Figure~\ref{fig:EM_Heis}(b) shows the staggered magnetization measured on the optimized state, which approaches to the extrapolated QMC results at larger bond dimensions. 

To obtain results for bond dimension $D>4$, we need to employ either the checkpoint technique~\ref{sec:checkpoint} or the fixed point iteration~\ref{sec:fixedpoint} to keep the memory budget low enough to fit into a single Nvidia P100 GPU card with $12$G memory. It is rather encouraging that with moderate effort one can reach the state-of-the-art performance in variational optimizing iPEPS~\cite{Corboz2016, Vanderstraeten2016}. The success is also a nontrivial demonstration that one can indeed stabilize reverse mode automatic differentiation for linear algebra operations appeared in scientific computation~\cite{Tamayo-Mendoza2018}. 

We note that the present approach applies as well to finite systems or problems with larger unit cells, more complex Hamiltonians~\cite{Corboz2011, Liao2017, Lee2018f, Haghshenas2018}, and more sophisticated contraction schemes with improved efficiency~\cite{Xie2017}, which is promising to deliver new physical results to quantum many-body problems.

\section{Discussions}\label{sec:discussions}
Computing the gradient via automatic differentiation significantly boosts the power of existing tensor network algorithms. Researchers can focus on the core tensor network contraction algorithms without worrying about the tedious gradient calculations. The computational complexity of automatic differentiation is the same as the forward contraction of the tensor networks. 

Besides greatly reducing human efforts, the automatic differentiation approach also computes a slightly different gradient than Refs.~\cite{Corboz2016, Vanderstraeten2016}. The present approach computes numerical exact gradient of an approximated energy density via automatic differentiation. While Refs.~\cite{Corboz2016, Vanderstraeten2016} first derive analytical expression of the energy gradient as infinite tensor networks, and then contract these networks approximately to obtain approximated gradient. Thus, the two approaches perform \emph{differentiate the approximation} and \emph{approximate the derivative} respectively~\cite{Baydin2018}. Other than the general recommendation of~Ref.~\cite{Baydin2018}, we find that differentiating through approximated tensor network contraction can be advantageous for infinite systems whose analytical derivative is complicated to derive and approximate. 

In this paper, we have focused on the high level applications of automatic differentiation that differentiates through the whole contraction algorithms for optimizing tensor networks and computing physical observables. The same techniques are also applicable to low level cases such as finding optimal truncation bases or variational transformation of tensor networks~\cite{Yang2017c}. Moreover, besides the optimization of the expected energy of quantum problems, the approach is also relevant to variational contraction of tensor networks~\cite{Haegeman2016, Vanderstraeten2018a}. We expect differentiable programming techniques will become an integrated part of the standard tensor network toolbox. 

A bonus of implementing tensor network programs using deep learning frameworks~\cite{Abadi2016, Paszke2017, jaxpaper, innes2018don} is that one can readily enjoy the GPU acceleration. The calculations of this work were done with a single GPU card. Pushing this line of research further, we envision that it will be rewarding to deploy tensor network algorithms on emerging specialized hardware in a larger scale.

Finally, it is useful to comment on the difference of automatic differentiation for tensor networks and neural networks. Typical neural network architectures do not involve sophisticated linear algebra operations. However, with the development of tensorized neural networks~\cite{Ionescu2015} and applications of various tensor networks to machine learning problems~\cite{novikov2015tensorizing, Stoudenmire2016b, Stoudenmire2018g, Han2018, Cheng2019, Stokes}, the boundary between the two classes of networks is blurred. 
Thus, results presented this paper would also be relevant to tensor network machine learning applications when one moves to more sophisticated contraction schemes. 

\section{Acknowledgment}
We are grateful to Pan Zhang, Zhi-Yuan Xie, Wei Li, Ling Wang, Dian Wu, Xiu-Zhe Luo, Shuo-Hui Li, Song Cheng, Jing Chen, Xi Dai, Anders Sandvik, and Frank Verstraete for useful discussions. We thank Philippe Corboz and Laurens Vanderstraeten for providing the reference data shown in Fig.~\ref{fig:EM_Heis}. The authors are supported by the National Key Research and Development Project of China Grant No.~2017YFA0302901 and No.~2016YFA0302400, the National Natural Science Foundation of China Grant No.~11888101 and No.~11774398, and the Strategic Priority Research Program of Chinese Academy of Sciences Grant No.~XDB28000000.

\bibliography{refs,manualrefs}

\begin{thebibliography}{103}%
\makeatletter
\providecommand \@ifxundefined [1]{%
 \@ifx{#1\undefined}
}%
\providecommand \@ifnum [1]{%
 \ifnum #1\expandafter \@firstoftwo
 \else \expandafter \@secondoftwo
 \fi
}%
\providecommand \@ifx [1]{%
 \ifx #1\expandafter \@firstoftwo
 \else \expandafter \@secondoftwo
 \fi
}%
\providecommand \natexlab [1]{#1}%
\providecommand \enquote  [1]{``#1''}%
\providecommand \bibnamefont  [1]{#1}%
\providecommand \bibfnamefont [1]{#1}%
\providecommand \citenamefont [1]{#1}%
\providecommand \href@noop [0]{\@secondoftwo}%
\providecommand \href [0]{\begingroup \@sanitize@url \@href}%
\providecommand \@href[1]{\@@startlink{#1}\@@href}%
\providecommand \@@href[1]{\endgroup#1\@@endlink}%
\providecommand \@sanitize@url [0]{\catcode `\\12\catcode `\$12\catcode
  `\&12\catcode `\#12\catcode `\^12\catcode `\_12\catcode `\%12\relax}%
\providecommand \@@startlink[1]{}%
\providecommand \@@endlink[0]{}%
\providecommand \url  [0]{\begingroup\@sanitize@url \@url }%
\providecommand \@url [1]{\endgroup\@href {#1}{\urlprefix }}%
\providecommand \urlprefix  [0]{URL }%
\providecommand \Eprint [0]{\href }%
\providecommand \doibase [0]{http://dx.doi.org/}%
\providecommand \selectlanguage [0]{\@gobble}%
\providecommand \bibinfo  [0]{\@secondoftwo}%
\providecommand \bibfield  [0]{\@secondoftwo}%
\providecommand \translation [1]{[#1]}%
\providecommand \BibitemOpen [0]{}%
\providecommand \bibitemStop [0]{}%
\providecommand \bibitemNoStop [0]{.\EOS\space}%
\providecommand \EOS [0]{\spacefactor3000\relax}%
\providecommand \BibitemShut  [1]{\csname bibitem#1\endcsname}%
\let\auto@bib@innerbib\@empty
\bibitem [{\citenamefont {Or\'us}(2014)}]{Orus2014}%
  \BibitemOpen
  \bibfield  {author} {\bibinfo {author} {\bibfnamefont {Rom\'an}\ \bibnamefont
  {Or\'us}},\ }\bibfield  {title} {\enquote {\bibinfo {title} {{A practical
  introduction to tensor networks: Matrix product states and projected
  entangled pair states}},}\ }\href {\doibase 10.1016/j.aop.2014.06.013}
  {\bibfield  {journal} {\bibinfo  {journal} {Ann. Phys. (N. Y).}\ }\textbf
  {\bibinfo {volume} {349}},\ \bibinfo {pages} {117--158} (\bibinfo {year}
  {2014})},\ \Eprint {http://arxiv.org/abs/1306.2164} {arXiv:1306.2164}
  \BibitemShut {NoStop}%
\bibitem [{\citenamefont {Haegeman}\ and\ \citenamefont
  {Verstraete}(2017)}]{Haegeman2016}%
  \BibitemOpen
  \bibfield  {author} {\bibinfo {author} {\bibfnamefont {Jutho}\ \bibnamefont
  {Haegeman}}\ and\ \bibinfo {author} {\bibfnamefont {Frank}\ \bibnamefont
  {Verstraete}},\ }\bibfield  {title} {\enquote {\bibinfo {title}
  {{Diagonalizing transfer matrices and matrix product operators: a medley of
  exact and computational methods}},}\ }\href {\doibase
  10.1146/annurev-conmatphys-031016-025507} {\bibfield  {journal} {\bibinfo
  {journal} {Annu. Rev. Condens. Matter Phys.}\ }\textbf {\bibinfo {volume}
  {8}},\ \bibinfo {pages} {355} (\bibinfo {year} {2017})},\ \Eprint
  {http://arxiv.org/abs/1611.08519} {arXiv:1611.08519} \BibitemShut {NoStop}%
\bibitem [{\citenamefont {Or\'us}(2018)}]{Orus2018}%
  \BibitemOpen
  \bibfield  {author} {\bibinfo {author} {\bibfnamefont {Rom\'an}\ \bibnamefont
  {Or\'us}},\ }\bibfield  {title} {\enquote {\bibinfo {title} {{Tensor networks
  for complex quantum systems}},}\ }\href {http://arxiv.org/abs/1812.04011}
  {\bibfield  {journal} {\bibinfo  {journal} {arXiv}\ } (\bibinfo {year}
  {2018})},\ \Eprint {http://arxiv.org/abs/1812.04011} {arXiv:1812.04011}
  \BibitemShut {NoStop}%
\bibitem [{\citenamefont {Markov}\ and\ \citenamefont {Shi}(2008)}]{treewidth}%
  \BibitemOpen
  \bibfield  {author} {\bibinfo {author} {\bibfnamefont {Igor}\ \bibnamefont
  {Markov}}\ and\ \bibinfo {author} {\bibfnamefont {Yaoyun}\ \bibnamefont
  {Shi}},\ }\bibfield  {title} {\enquote {\bibinfo {title} {{Simulating quantum
  computation by contracting tensor networks}},}\ }\href
  {https://epubs.siam.org/doi/10.1137/050644756} {\bibfield  {journal}
  {\bibinfo  {journal} {SIAM J. Comput.}\ }\textbf {\bibinfo {volume} {38}},\
  \bibinfo {pages} {963} (\bibinfo {year} {2008})}\BibitemShut {NoStop}%
\bibitem [{\citenamefont {Arad}\ and\ \citenamefont {Landau}(2010)}]{Arad2010}%
  \BibitemOpen
  \bibfield  {author} {\bibinfo {author} {\bibfnamefont {Itai}\ \bibnamefont
  {Arad}}\ and\ \bibinfo {author} {\bibfnamefont {Zeph}\ \bibnamefont
  {Landau}},\ }\bibfield  {title} {\enquote {\bibinfo {title} {{Quantum
  computation and the evaluation of tensor networks}},}\ }\href {\doibase
  doi.org/10.1137/080739379} {\bibfield  {journal} {\bibinfo  {journal} {SIAM
  J. Comput.}\ }\textbf {\bibinfo {volume} {39}},\ \bibinfo {pages} {3089}
  (\bibinfo {year} {2010})},\ \Eprint {http://arxiv.org/abs/0805.0040v3}
  {arXiv:0805.0040v3} \BibitemShut {NoStop}%
\bibitem [{\citenamefont {Kim}\ and\ \citenamefont {Swingle}(2017)}]{Kim2017a}%
  \BibitemOpen
  \bibfield  {author} {\bibinfo {author} {\bibfnamefont {Isaac~H.}\
  \bibnamefont {Kim}}\ and\ \bibinfo {author} {\bibfnamefont {Brian}\
  \bibnamefont {Swingle}},\ }\bibfield  {title} {\enquote {\bibinfo {title}
  {{Robust entanglement renormalization on a noisy quantum computer}},}\ }\href
  {http://arxiv.org/abs/1711.07500} {\bibfield  {journal} {\bibinfo  {journal}
  {arXiv}\ } (\bibinfo {year} {2017})},\ \Eprint
  {http://arxiv.org/abs/1711.07500} {arXiv:1711.07500} \BibitemShut {NoStop}%
\bibitem [{\citenamefont {Huggins}\ \emph {et~al.}(2019)\citenamefont
  {Huggins}, \citenamefont {Patel}, \citenamefont {Whaley},\ and\ \citenamefont
  {Stoudenmire}}]{Huggins2019}%
  \BibitemOpen
  \bibfield  {author} {\bibinfo {author} {\bibfnamefont {William}\ \bibnamefont
  {Huggins}}, \bibinfo {author} {\bibfnamefont {Piyush}\ \bibnamefont {Patel}},
  \bibinfo {author} {\bibfnamefont {K.~Birgitta}\ \bibnamefont {Whaley}}, \
  and\ \bibinfo {author} {\bibfnamefont {E.~Miles}\ \bibnamefont
  {Stoudenmire}},\ }\bibfield  {title} {\enquote {\bibinfo {title} {{Towards
  Quantum Machine Learning with Tensor Networks}},}\ }\href {\doibase
  10.1021/ac3004105} {\bibfield  {journal} {\bibinfo  {journal} {Quantum Sci.
  Technol.}\ }\textbf {\bibinfo {volume} {4}},\ \bibinfo {pages} {024001}
  (\bibinfo {year} {2019})},\ \Eprint {http://arxiv.org/abs/1803.11537}
  {arXiv:1803.11537} \BibitemShut {NoStop}%
\bibitem [{\citenamefont {Ferris}\ and\ \citenamefont
  {Poulin}(2014)}]{Ferris2014a}%
  \BibitemOpen
  \bibfield  {author} {\bibinfo {author} {\bibfnamefont {Andrew~J}\
  \bibnamefont {Ferris}}\ and\ \bibinfo {author} {\bibfnamefont {David}\
  \bibnamefont {Poulin}},\ }\bibfield  {title} {\enquote {\bibinfo {title}
  {{Tensor networks and quantum error correction}},}\ }\href {\doibase
  10.1103/PhysRevLett.113.030501} {\bibfield  {journal} {\bibinfo  {journal}
  {Phys. Rev. Lett.}\ }\textbf {\bibinfo {volume} {113}},\ \bibinfo {pages}
  {030501} (\bibinfo {year} {2014})}\BibitemShut {NoStop}%
\bibitem [{\citenamefont {Bravyi}\ \emph {et~al.}(2014)\citenamefont {Bravyi},
  \citenamefont {Suchara},\ and\ \citenamefont {Vargo}}]{Bravyi2014}%
  \BibitemOpen
  \bibfield  {author} {\bibinfo {author} {\bibfnamefont {Sergey}\ \bibnamefont
  {Bravyi}}, \bibinfo {author} {\bibfnamefont {Martin}\ \bibnamefont
  {Suchara}}, \ and\ \bibinfo {author} {\bibfnamefont {Alexander}\ \bibnamefont
  {Vargo}},\ }\bibfield  {title} {\enquote {\bibinfo {title} {{Efficient
  algorithms for maximum likelihood decoding in the surface code}},}\ }\href
  {\doibase 10.1103/PhysRevA.90.032326} {\bibfield  {journal} {\bibinfo
  {journal} {Phys. Rev. A}\ }\textbf {\bibinfo {volume} {90}},\ \bibinfo
  {pages} {032326} (\bibinfo {year} {2014})}\BibitemShut {NoStop}%
\bibitem [{\citenamefont {Stoudenmire}\ and\ \citenamefont
  {Schwab}(2016)}]{Stoudenmire2016b}%
  \BibitemOpen
  \bibfield  {author} {\bibinfo {author} {\bibfnamefont {E.~Miles}\
  \bibnamefont {Stoudenmire}}\ and\ \bibinfo {author} {\bibfnamefont
  {David~J.}\ \bibnamefont {Schwab}},\ }\bibfield  {title} {\enquote {\bibinfo
  {title} {{Supervised Learning with Tensor Networks}},}\ }\href
  {https://papers.nips.cc/paper/6211-supervised-learning-with-tensor-networksAdvances
  in Neural Information Processing Systems 29 (NIPS
  2016){\%}0Ahttp://arxiv.org/abs/1605.05775} {\bibfield  {journal} {\bibinfo
  {journal} {Adv. Neural Inf. Process. Syst. 29 (NIPS 2016)}\ ,\ \bibinfo
  {pages} {4799--4807}} (\bibinfo {year} {2016})},\ \Eprint
  {http://arxiv.org/abs/1605.05775} {arXiv:1605.05775} \BibitemShut {NoStop}%
\bibitem [{\citenamefont {Stoudenmire}(2018)}]{Stoudenmire2018g}%
  \BibitemOpen
  \bibfield  {author} {\bibinfo {author} {\bibfnamefont {E~Miles}\ \bibnamefont
  {Stoudenmire}},\ }\bibfield  {title} {\enquote {\bibinfo {title} {{Learning
  relevant features of data with multi-scale tensor networks}},}\ }\href
  {\doibase 10.1088/2058-9565/aaba1a} {\bibfield  {journal} {\bibinfo
  {journal} {Quantum Sci. Technol.}\ }\textbf {\bibinfo {volume} {3}},\
  \bibinfo {pages} {034003} (\bibinfo {year} {2018})}\BibitemShut {NoStop}%
\bibitem [{\citenamefont {Han}\ \emph {et~al.}(2018)\citenamefont {Han},
  \citenamefont {Wang}, \citenamefont {Fan}, \citenamefont {Wang},\ and\
  \citenamefont {Zhang}}]{Han2018}%
  \BibitemOpen
  \bibfield  {author} {\bibinfo {author} {\bibfnamefont {Zhao-Yu}\ \bibnamefont
  {Han}}, \bibinfo {author} {\bibfnamefont {Jun}\ \bibnamefont {Wang}},
  \bibinfo {author} {\bibfnamefont {Heng}\ \bibnamefont {Fan}}, \bibinfo
  {author} {\bibfnamefont {Lei}\ \bibnamefont {Wang}}, \ and\ \bibinfo {author}
  {\bibfnamefont {Pan}\ \bibnamefont {Zhang}},\ }\bibfield  {title} {\enquote
  {\bibinfo {title} {{Unsupervised Generative Modeling Using Matrix Product
  States}},}\ }\href {\doibase 10.1103/PhysRevX.8.031012} {\bibfield  {journal}
  {\bibinfo  {journal} {Phys. Rev. X}\ }\textbf {\bibinfo {volume} {8}},\
  \bibinfo {pages} {31012} (\bibinfo {year} {2018})}\BibitemShut {NoStop}%
\bibitem [{\citenamefont {Cheng}\ \emph {et~al.}(2019)\citenamefont {Cheng},
  \citenamefont {Wang}, \citenamefont {Xiang},\ and\ \citenamefont
  {Zhang}}]{Cheng2019}%
  \BibitemOpen
  \bibfield  {author} {\bibinfo {author} {\bibfnamefont {Song}\ \bibnamefont
  {Cheng}}, \bibinfo {author} {\bibfnamefont {Lei}\ \bibnamefont {Wang}},
  \bibinfo {author} {\bibfnamefont {Tao}\ \bibnamefont {Xiang}}, \ and\
  \bibinfo {author} {\bibfnamefont {Pan}\ \bibnamefont {Zhang}},\ }\bibfield
  {title} {\enquote {\bibinfo {title} {Tree tensor networks for generative
  modeling},}\ }\href {\doibase 10.1103/PhysRevB.99.155131} {\bibfield
  {journal} {\bibinfo  {journal} {Phys. Rev. B}\ }\textbf {\bibinfo {volume}
  {99}},\ \bibinfo {pages} {155131} (\bibinfo {year} {2019})}\BibitemShut
  {NoStop}%
\bibitem [{\citenamefont {Stokes}\ and\ \citenamefont
  {Terilla}(2019)}]{Stokes}%
  \BibitemOpen
  \bibfield  {author} {\bibinfo {author} {\bibfnamefont {James}\ \bibnamefont
  {Stokes}}\ and\ \bibinfo {author} {\bibfnamefont {John}\ \bibnamefont
  {Terilla}},\ }\bibfield  {title} {\enquote {\bibinfo {title} {{Probabilistic
  Modeling With Matrix Product States}},}\ }\href
  {https://arxiv.org/pdf/1902.06888.pdf} {\bibfield  {journal} {\bibinfo
  {journal} {arXiv}\ } (\bibinfo {year} {2019})},\ \Eprint
  {http://arxiv.org/abs/1902.06888v1} {arXiv:1902.06888v1} \BibitemShut
  {NoStop}%
\bibitem [{\citenamefont {Gallego}\ and\ \citenamefont
  {Or\'us}(2017)}]{Gallego2017}%
  \BibitemOpen
  \bibfield  {author} {\bibinfo {author} {\bibfnamefont {Angel~J.}\
  \bibnamefont {Gallego}}\ and\ \bibinfo {author} {\bibfnamefont {Rom\'an}\
  \bibnamefont {Or\'us}},\ }\bibfield  {title} {\enquote {\bibinfo {title}
  {{Language Design and Renormalization}},}\ }\href
  {http://arxiv.org/abs/1708.01525} {\bibfield  {journal} {\bibinfo  {journal}
  {arXiv}\ } (\bibinfo {year} {2017})},\ \Eprint
  {http://arxiv.org/abs/1708.01525} {arXiv:1708.01525} \BibitemShut {NoStop}%
\bibitem [{\citenamefont {Pestun}\ and\ \citenamefont
  {Vlassopoulos}(2017)}]{Pestun2017a}%
  \BibitemOpen
  \bibfield  {author} {\bibinfo {author} {\bibfnamefont {Vasily}\ \bibnamefont
  {Pestun}}\ and\ \bibinfo {author} {\bibfnamefont {Yiannis}\ \bibnamefont
  {Vlassopoulos}},\ }\bibfield  {title} {\enquote {\bibinfo {title} {{Tensor
  network language model}},}\ }\href {http://arxiv.org/abs/1710.10248}
  {\bibfield  {journal} {\bibinfo  {journal} {arXiv}\ } (\bibinfo {year}
  {2017})},\ \Eprint {http://arxiv.org/abs/1710.10248} {arXiv:1710.10248}
  \BibitemShut {NoStop}%
\bibitem [{\citenamefont {Verstraete}\ and\ \citenamefont
  {Cirac}(2010)}]{Verstraete2010a}%
  \BibitemOpen
  \bibfield  {author} {\bibinfo {author} {\bibfnamefont {F.}~\bibnamefont
  {Verstraete}}\ and\ \bibinfo {author} {\bibfnamefont {J.~I.}\ \bibnamefont
  {Cirac}},\ }\bibfield  {title} {\enquote {\bibinfo {title} {{Continuous
  matrix product states for quantum fields}},}\ }\href {\doibase
  10.1103/PhysRevLett.104.190405} {\bibfield  {journal} {\bibinfo  {journal}
  {Phys. Rev. Lett.}\ }\textbf {\bibinfo {volume} {104}},\ \bibinfo {pages}
  {190405} (\bibinfo {year} {2010})}\BibitemShut {NoStop}%
\bibitem [{\citenamefont {Haegeman}\ \emph {et~al.}(2013)\citenamefont
  {Haegeman}, \citenamefont {Osborne}, \citenamefont {Verschelde},\ and\
  \citenamefont {Verstraete}}]{Haegeman2013}%
  \BibitemOpen
  \bibfield  {author} {\bibinfo {author} {\bibfnamefont {Jutho}\ \bibnamefont
  {Haegeman}}, \bibinfo {author} {\bibfnamefont {Tobias~J}\ \bibnamefont
  {Osborne}}, \bibinfo {author} {\bibfnamefont {Henri}\ \bibnamefont
  {Verschelde}}, \ and\ \bibinfo {author} {\bibfnamefont {Frank}\ \bibnamefont
  {Verstraete}},\ }\bibfield  {title} {\enquote {\bibinfo {title}
  {{Entanglement renormalization for quantum fields}},}\ }\href {\doibase
  10.1103/PhysRevLett.110.100402} {\bibfield  {journal} {\bibinfo  {journal}
  {Phys. Rev. Lett.}\ }\textbf {\bibinfo {volume} {110}},\ \bibinfo {pages}
  {100402} (\bibinfo {year} {2013})},\ \Eprint {http://arxiv.org/abs/1102.5524}
  {arXiv:1102.5524} \BibitemShut {NoStop}%
\bibitem [{\citenamefont {Hu}\ \emph {et~al.}(2018)\citenamefont {Hu},
  \citenamefont {Franco-Rubio},\ and\ \citenamefont {Vidal}}]{Hu2018}%
  \BibitemOpen
  \bibfield  {author} {\bibinfo {author} {\bibfnamefont {Qi}~\bibnamefont
  {Hu}}, \bibinfo {author} {\bibfnamefont {Adrian}\ \bibnamefont
  {Franco-Rubio}}, \ and\ \bibinfo {author} {\bibfnamefont {Guifre}\
  \bibnamefont {Vidal}},\ }\bibfield  {title} {\enquote {\bibinfo {title}
  {{Continuous tensor network renormalization for quantum fields}},}\ }\href
  {http://arxiv.org/abs/1809.05176} {\bibfield  {journal} {\bibinfo  {journal}
  {arXiv}\ } (\bibinfo {year} {2018})},\ \Eprint
  {http://arxiv.org/abs/1809.05176} {arXiv:1809.05176} \BibitemShut {NoStop}%
\bibitem [{\citenamefont {Tilloy}\ and\ \citenamefont
  {Cirac}(2019)}]{Tilloy2018}%
  \BibitemOpen
  \bibfield  {author} {\bibinfo {author} {\bibfnamefont {Antoine}\ \bibnamefont
  {Tilloy}}\ and\ \bibinfo {author} {\bibfnamefont {J.~Ignacio}\ \bibnamefont
  {Cirac}},\ }\bibfield  {title} {\enquote {\bibinfo {title} {Continuous tensor
  network states for quantum fields},}\ }\href {\doibase
  10.1103/PhysRevX.9.021040} {\bibfield  {journal} {\bibinfo  {journal} {Phys.
  Rev. X}\ }\textbf {\bibinfo {volume} {9}},\ \bibinfo {pages} {021040}
  (\bibinfo {year} {2019})}\BibitemShut {NoStop}%
\bibitem [{\citenamefont {Swingle}(2012)}]{Swingle2012}%
  \BibitemOpen
  \bibfield  {author} {\bibinfo {author} {\bibfnamefont {Brian}\ \bibnamefont
  {Swingle}},\ }\bibfield  {title} {\enquote {\bibinfo {title} {{Entanglement
  renormalization and holography}},}\ }\href {\doibase
  10.1103/PhysRevD.86.065007} {\bibfield  {journal} {\bibinfo  {journal} {Phys.
  Rev. D}\ }\textbf {\bibinfo {volume} {86}},\ \bibinfo {pages} {065007}
  (\bibinfo {year} {2012})}\BibitemShut {NoStop}%
\bibitem [{\citenamefont {Hayden}\ \emph {et~al.}(2016)\citenamefont {Hayden},
  \citenamefont {Nezami}, \citenamefont {Qi}, \citenamefont {Thomas},
  \citenamefont {Walter},\ and\ \citenamefont {Yang}}]{Hayden2016}%
  \BibitemOpen
  \bibfield  {author} {\bibinfo {author} {\bibfnamefont {Patrick}\ \bibnamefont
  {Hayden}}, \bibinfo {author} {\bibfnamefont {Sepehr}\ \bibnamefont {Nezami}},
  \bibinfo {author} {\bibfnamefont {Xiao~Liang}\ \bibnamefont {Qi}}, \bibinfo
  {author} {\bibfnamefont {Nathaniel}\ \bibnamefont {Thomas}}, \bibinfo
  {author} {\bibfnamefont {Michael}\ \bibnamefont {Walter}}, \ and\ \bibinfo
  {author} {\bibfnamefont {Zhao}\ \bibnamefont {Yang}},\ }\bibfield  {title}
  {\enquote {\bibinfo {title} {{Holographic duality from random tensor
  networks}},}\ }\href {\doibase 10.1007/JHEP11(2016)009} {\bibfield  {journal}
  {\bibinfo  {journal} {J. High Energy Phys.}\ }\textbf {\bibinfo {volume}
  {11}},\ \bibinfo {pages} {009} (\bibinfo {year} {2016})},\ \Eprint
  {http://arxiv.org/abs/1601.01694} {arXiv:1601.01694} \BibitemShut {NoStop}%
\bibitem [{\citenamefont {White}(1992)}]{White1992}%
  \BibitemOpen
  \bibfield  {author} {\bibinfo {author} {\bibfnamefont {Steven~R.}\
  \bibnamefont {White}},\ }\bibfield  {title} {\enquote {\bibinfo {title}
  {{Density matrix formulation for quantum renormalization groups}},}\ }\href
  {\doibase 10.1103/PhysRevLett.69.2863} {\bibfield  {journal} {\bibinfo
  {journal} {Phys. Rev. Lett.}\ }\textbf {\bibinfo {volume} {69}},\ \bibinfo
  {pages} {2863} (\bibinfo {year} {1992})}\BibitemShut {NoStop}%
\bibitem [{\citenamefont {Vidal}(2006)}]{Vidal2006}%
  \BibitemOpen
  \bibfield  {author} {\bibinfo {author} {\bibfnamefont {G.}~\bibnamefont
  {Vidal}},\ }\bibfield  {title} {\enquote {\bibinfo {title} {{Classical
  simulation of infinite-size quantum lattice systems in one spatial
  dimension}},}\ }\href {\doibase 10.1103/PhysRevLett.98.070201} {\bibfield
  {journal} {\bibinfo  {journal} {Phys. Rev. Lett.}\ }\textbf {\bibinfo
  {volume} {98}},\ \bibinfo {pages} {070201} (\bibinfo {year} {2006})},\
  \Eprint {http://arxiv.org/abs/0605597} {arXiv:0605597 [cond-mat]}
  \BibitemShut {NoStop}%
\bibitem [{\citenamefont {Schollw{\"{o}}ck}(2011)}]{Schollwock2011}%
  \BibitemOpen
  \bibfield  {author} {\bibinfo {author} {\bibfnamefont {Ulrich}\ \bibnamefont
  {Schollw{\"{o}}ck}},\ }\bibfield  {title} {\enquote {\bibinfo {title} {{The
  density-matrix renormalization group in the age of matrix product states}},}\
  }\href {\doibase 10.1016/j.aop.2010.09.012} {\bibfield  {journal} {\bibinfo
  {journal} {Ann. Phys. (N. Y).}\ }\textbf {\bibinfo {volume} {326}},\ \bibinfo
  {pages} {96--192} (\bibinfo {year} {2011})},\ \Eprint
  {http://arxiv.org/abs/1008.3477} {arXiv:1008.3477} \BibitemShut {NoStop}%
\bibitem [{\citenamefont {Stoudenmire}\ and\ \citenamefont
  {White}(2011)}]{Stoudenmire2011}%
  \BibitemOpen
  \bibfield  {author} {\bibinfo {author} {\bibfnamefont {E.~M.}\ \bibnamefont
  {Stoudenmire}}\ and\ \bibinfo {author} {\bibfnamefont {Steven~R.}\
  \bibnamefont {White}},\ }\bibfield  {title} {\enquote {\bibinfo {title}
  {{Studying Two Dimensional Systems With the Density Matrix Renormalization
  Group}},}\ }\href {\doibase 10.1146/annurev-conmatphys-020911-125018}
  {\bibfield  {journal} {\bibinfo  {journal} {Annu. Rev. Condens. Matter
  Phys.}\ }\textbf {\bibinfo {volume} {3}} (\bibinfo {year} {2011}),\
  10.1146/annurev-conmatphys-020911-125018},\ \Eprint
  {http://arxiv.org/abs/1105.1374} {arXiv:1105.1374} \BibitemShut {NoStop}%
\bibitem [{\citenamefont {Landau}\ \emph {et~al.}(2015)\citenamefont {Landau},
  \citenamefont {Vazirani},\ and\ \citenamefont {Vidick}}]{Landau2015}%
  \BibitemOpen
  \bibfield  {author} {\bibinfo {author} {\bibfnamefont {Zeph}\ \bibnamefont
  {Landau}}, \bibinfo {author} {\bibfnamefont {Umesh}\ \bibnamefont
  {Vazirani}}, \ and\ \bibinfo {author} {\bibfnamefont {Thomas}\ \bibnamefont
  {Vidick}},\ }\bibfield  {title} {\enquote {\bibinfo {title} {{A polynomial
  time algorithm for the ground state of one-dimensional gapped local
  Hamiltonians}},}\ }\href {\doibase 10.1038/nphys3345} {\bibfield  {journal}
  {\bibinfo  {journal} {Nat. Phys.}\ }\textbf {\bibinfo {volume} {11}},\
  \bibinfo {pages} {566--569} (\bibinfo {year} {2015})}\BibitemShut {NoStop}%
\bibitem [{\citenamefont {Zauner-Stauber}\ \emph {et~al.}(2018)\citenamefont
  {Zauner-Stauber}, \citenamefont {Vanderstraeten}, \citenamefont {Fishman},
  \citenamefont {Verstraete},\ and\ \citenamefont
  {Haegeman}}]{Zauner-Stauber2018}%
  \BibitemOpen
  \bibfield  {author} {\bibinfo {author} {\bibfnamefont {V.}~\bibnamefont
  {Zauner-Stauber}}, \bibinfo {author} {\bibfnamefont {L}~\bibnamefont
  {Vanderstraeten}}, \bibinfo {author} {\bibfnamefont {M~T}\ \bibnamefont
  {Fishman}}, \bibinfo {author} {\bibfnamefont {F}~\bibnamefont {Verstraete}},
  \ and\ \bibinfo {author} {\bibfnamefont {J}~\bibnamefont {Haegeman}},\
  }\bibfield  {title} {\enquote {\bibinfo {title} {{Variational optimization
  algorithms for uniform matrix product states}},}\ }\href {\doibase
  10.1103/PhysRevB.97.045145} {\bibfield  {journal} {\bibinfo  {journal} {Phys.
  Rev. B}\ }\textbf {\bibinfo {volume} {97}},\ \bibinfo {pages} {045145}
  (\bibinfo {year} {2018})},\ \Eprint {http://arxiv.org/abs/1701.07035v1}
  {arXiv:1701.07035v1} \BibitemShut {NoStop}%
\bibitem [{\citenamefont {Jiang}\ \emph {et~al.}(2008)\citenamefont {Jiang},
  \citenamefont {Weng},\ and\ \citenamefont {Xiang}}]{Jiang2008}%
  \BibitemOpen
  \bibfield  {author} {\bibinfo {author} {\bibfnamefont {H.~C.}\ \bibnamefont
  {Jiang}}, \bibinfo {author} {\bibfnamefont {Z.~Y.}\ \bibnamefont {Weng}}, \
  and\ \bibinfo {author} {\bibfnamefont {T.}~\bibnamefont {Xiang}},\ }\bibfield
   {title} {\enquote {\bibinfo {title} {{Accurate Determination of Tensor
  Network State of Quantum Lattice Models in Two Dimensions}},}\ }\href
  {\doibase 10.1103/PhysRevLett.101.090603} {\bibfield  {journal} {\bibinfo
  {journal} {Phys. Rev. Lett.}\ }\textbf {\bibinfo {volume} {101}},\ \bibinfo
  {pages} {090603} (\bibinfo {year} {2008})}\BibitemShut {NoStop}%
\bibitem [{\citenamefont {Jordan}\ \emph {et~al.}(2008)\citenamefont {Jordan},
  \citenamefont {Or{\'{u}}s}, \citenamefont {Vidal}, \citenamefont
  {Verstraete},\ and\ \citenamefont {Cirac}}]{Jordan2008}%
  \BibitemOpen
  \bibfield  {author} {\bibinfo {author} {\bibfnamefont {J.}~\bibnamefont
  {Jordan}}, \bibinfo {author} {\bibfnamefont {R.}~\bibnamefont {Or{\'{u}}s}},
  \bibinfo {author} {\bibfnamefont {G.}~\bibnamefont {Vidal}}, \bibinfo
  {author} {\bibfnamefont {F.}~\bibnamefont {Verstraete}}, \ and\ \bibinfo
  {author} {\bibfnamefont {J.~I.}\ \bibnamefont {Cirac}},\ }\bibfield  {title}
  {\enquote {\bibinfo {title} {{Classical simulation of infinite-size quantum
  lattice systems in two spatial dimensions}},}\ }\href {\doibase
  10.1103/PhysRevLett.101.250602} {\bibfield  {journal} {\bibinfo  {journal}
  {Phys. Rev. Lett.}\ }\textbf {\bibinfo {volume} {101}},\ \bibinfo {pages}
  {250602} (\bibinfo {year} {2008})},\ \Eprint {http://arxiv.org/abs/0703788}
  {arXiv:0703788 [cond-mat]} \BibitemShut {NoStop}%
\bibitem [{\citenamefont {Corboz}\ \emph {et~al.}(2010)\citenamefont {Corboz},
  \citenamefont {Or\'us}, \citenamefont {Bauer},\ and\ \citenamefont
  {Vidal}}]{Corboz2010}%
  \BibitemOpen
  \bibfield  {author} {\bibinfo {author} {\bibfnamefont {Philippe}\
  \bibnamefont {Corboz}}, \bibinfo {author} {\bibfnamefont {Rom\'an}\
  \bibnamefont {Or\'us}}, \bibinfo {author} {\bibfnamefont {Bela}\ \bibnamefont
  {Bauer}}, \ and\ \bibinfo {author} {\bibfnamefont {Guifr\'e}\ \bibnamefont
  {Vidal}},\ }\bibfield  {title} {\enquote {\bibinfo {title} {{Simulation of
  strongly correlated fermions in two spatial dimensions with fermionic
  projected entangled-pair states}},}\ }\href {\doibase
  10.1103/PhysRevB.81.165104} {\bibfield  {journal} {\bibinfo  {journal} {Phys.
  Rev. B}\ }\textbf {\bibinfo {volume} {81}},\ \bibinfo {pages} {165104}
  (\bibinfo {year} {2010})},\ \Eprint {http://arxiv.org/abs/0912.0646}
  {arXiv:0912.0646} \BibitemShut {NoStop}%
\bibitem [{\citenamefont {Zhao}\ \emph {et~al.}(2010)\citenamefont {Zhao},
  \citenamefont {Xie}, \citenamefont {Chen}, \citenamefont {Wei}, \citenamefont
  {Cai},\ and\ \citenamefont {Xiang}}]{Zhao2010}%
  \BibitemOpen
  \bibfield  {author} {\bibinfo {author} {\bibfnamefont {H.~H.}\ \bibnamefont
  {Zhao}}, \bibinfo {author} {\bibfnamefont {Z.~Y.}\ \bibnamefont {Xie}},
  \bibinfo {author} {\bibfnamefont {Q.~N.}\ \bibnamefont {Chen}}, \bibinfo
  {author} {\bibfnamefont {Z.~C.}\ \bibnamefont {Wei}}, \bibinfo {author}
  {\bibfnamefont {J.~W.}\ \bibnamefont {Cai}}, \ and\ \bibinfo {author}
  {\bibfnamefont {T.}~\bibnamefont {Xiang}},\ }\bibfield  {title} {\enquote
  {\bibinfo {title} {{Renormalization of tensor-network states}},}\ }\href
  {\doibase 10.1103/PhysRevB.81.174411} {\bibfield  {journal} {\bibinfo
  {journal} {Phys. Rev. B}\ }\textbf {\bibinfo {volume} {81}},\ \bibinfo
  {pages} {174411} (\bibinfo {year} {2010})},\ \Eprint
  {http://arxiv.org/abs/1002.1405} {arXiv:1002.1405} \BibitemShut {NoStop}%
\bibitem [{\citenamefont {Phien}\ \emph {et~al.}(2015)\citenamefont {Phien},
  \citenamefont {Bengua}, \citenamefont {Tuan}, \citenamefont {Corboz},\ and\
  \citenamefont {Or{\'{u}}s}}]{Phien2015}%
  \BibitemOpen
  \bibfield  {author} {\bibinfo {author} {\bibfnamefont {Ho~N.}\ \bibnamefont
  {Phien}}, \bibinfo {author} {\bibfnamefont {Johann~A.}\ \bibnamefont
  {Bengua}}, \bibinfo {author} {\bibfnamefont {Hoang~D.}\ \bibnamefont {Tuan}},
  \bibinfo {author} {\bibfnamefont {Philippe}\ \bibnamefont {Corboz}}, \ and\
  \bibinfo {author} {\bibfnamefont {Rom{\'{a}}n}\ \bibnamefont {Or{\'{u}}s}},\
  }\bibfield  {title} {\enquote {\bibinfo {title} {{Infinite projected
  entangled pair states algorithm improved: Fast full update and gauge
  fixing}},}\ }\href {\doibase 10.1103/PhysRevB.92.035142} {\bibfield
  {journal} {\bibinfo  {journal} {Phys. Rev. B}\ }\textbf {\bibinfo {volume}
  {92}},\ \bibinfo {pages} {035142} (\bibinfo {year} {2015})},\ \Eprint
  {http://arxiv.org/abs/1503.05345} {arXiv:1503.05345} \BibitemShut {NoStop}%
\bibitem [{\citenamefont {Corboz}(2016)}]{Corboz2016}%
  \BibitemOpen
  \bibfield  {author} {\bibinfo {author} {\bibfnamefont {Philippe}\
  \bibnamefont {Corboz}},\ }\bibfield  {title} {\enquote {\bibinfo {title}
  {{Variational optimization with infinite projected entangled-pair states}},}\
  }\href {\doibase 10.1103/PhysRevB.94.035133} {\bibfield  {journal} {\bibinfo
  {journal} {Phys. Rev. B}\ }\textbf {\bibinfo {volume} {94}},\ \bibinfo
  {pages} {035133} (\bibinfo {year} {2016})},\ \Eprint
  {http://arxiv.org/abs/1605.03006} {arXiv:1605.03006} \BibitemShut {NoStop}%
\bibitem [{\citenamefont {Vanderstraeten}\ \emph {et~al.}(2016)\citenamefont
  {Vanderstraeten}, \citenamefont {Haegeman}, \citenamefont {Corboz},\ and\
  \citenamefont {Verstraete}}]{Vanderstraeten2016}%
  \BibitemOpen
  \bibfield  {author} {\bibinfo {author} {\bibfnamefont {Laurens}\ \bibnamefont
  {Vanderstraeten}}, \bibinfo {author} {\bibfnamefont {Jutho}\ \bibnamefont
  {Haegeman}}, \bibinfo {author} {\bibfnamefont {Philippe}\ \bibnamefont
  {Corboz}}, \ and\ \bibinfo {author} {\bibfnamefont {Frank}\ \bibnamefont
  {Verstraete}},\ }\bibfield  {title} {\enquote {\bibinfo {title} {{Gradient
  methods for variational optimization of projected entangled-pair states}},}\
  }\href {\doibase 10.1103/PhysRevB.94.155123} {\bibfield  {journal} {\bibinfo
  {journal} {Phys. Rev. B}\ }\textbf {\bibinfo {volume} {94}},\ \bibinfo
  {pages} {155123} (\bibinfo {year} {2016})},\ \Eprint
  {http://arxiv.org/abs/1606.09170} {arXiv:1606.09170} \BibitemShut {NoStop}%
\bibitem [{\citenamefont {Poilblanc}\ and\ \citenamefont
  {Mambrini}(2017)}]{Poilblanc2017}%
  \BibitemOpen
  \bibfield  {author} {\bibinfo {author} {\bibfnamefont {Didier}\ \bibnamefont
  {Poilblanc}}\ and\ \bibinfo {author} {\bibfnamefont {Matthieu}\ \bibnamefont
  {Mambrini}},\ }\bibfield  {title} {\enquote {\bibinfo {title} {{Quantum
  critical phase with infinite projected entangled paired states}},}\ }\href
  {\doibase https://doi.org/10.1103/PhysRevB.96.014414} {\bibfield  {journal}
  {\bibinfo  {journal} {Phys. Rev. B}\ }\textbf {\bibinfo {volume} {96}},\
  \bibinfo {pages} {014414} (\bibinfo {year} {2017})}\BibitemShut {NoStop}%
\bibitem [{\citenamefont {Chen}\ \emph {et~al.}(2018)\citenamefont {Chen},
  \citenamefont {Vanderstraeten}, \citenamefont {Capponi},\ and\ \citenamefont
  {Poilblanc}}]{Chen2018k}%
  \BibitemOpen
  \bibfield  {author} {\bibinfo {author} {\bibfnamefont {Ji-Yao}\ \bibnamefont
  {Chen}}, \bibinfo {author} {\bibfnamefont {Laurens}\ \bibnamefont
  {Vanderstraeten}}, \bibinfo {author} {\bibfnamefont {Sylvain}\ \bibnamefont
  {Capponi}}, \ and\ \bibinfo {author} {\bibfnamefont {Didier}\ \bibnamefont
  {Poilblanc}},\ }\bibfield  {title} {\enquote {\bibinfo {title} {{Non-Abelian
  chiral spin liquid in a quantum antiferromagnet revealed by an iPEPS
  study}},}\ }\href {\doibase 10.1103/PhysRevB.98.184409} {\bibfield  {journal}
  {\bibinfo  {journal} {Phys. Rev. B}\ }\textbf {\bibinfo {volume} {98}},\
  \bibinfo {pages} {184409} (\bibinfo {year} {2018})},\ \Eprint
  {http://arxiv.org/abs/1807.04385} {arXiv:1807.04385} \BibitemShut {NoStop}%
\bibitem [{\citenamefont {Wang}\ \emph {et~al.}(2011)\citenamefont {Wang},
  \citenamefont {Kao},\ and\ \citenamefont {Sandvik}}]{Wang2011}%
  \BibitemOpen
  \bibfield  {author} {\bibinfo {author} {\bibfnamefont {Ling}\ \bibnamefont
  {Wang}}, \bibinfo {author} {\bibfnamefont {Ying-Jer}\ \bibnamefont {Kao}}, \
  and\ \bibinfo {author} {\bibfnamefont {Anders~W.}\ \bibnamefont {Sandvik}},\
  }\bibfield  {title} {\enquote {\bibinfo {title} {{Plaquette renormalization
  scheme for tensor network states}},}\ }\href {\doibase
  10.1103/PhysRevE.83.056703} {\bibfield  {journal} {\bibinfo  {journal} {Phys.
  Rev. E}\ }\textbf {\bibinfo {volume} {83}},\ \bibinfo {pages} {056703}
  (\bibinfo {year} {2011})},\ \Eprint {http://arxiv.org/abs/0901.0214}
  {arXiv:0901.0214} \BibitemShut {NoStop}%
\bibitem [{Git()}]{Github}%
  \BibitemOpen
  \href@noop {} {}\bibinfo {note} {See
  \href{https://github.com/wangleiphy/tensorgrad}{https://github.com/wangleiphy/tensorgrad}
  for code implementation in PyTorch}\BibitemShut {NoStop}%
\bibitem [{\citenamefont {Bartholomew-Biggs}\ \emph {et~al.}(2000)\citenamefont
  {Bartholomew-Biggs}, \citenamefont {Brown}, \citenamefont {Christianson},\
  and\ \citenamefont {Dixon}}]{Bartholomew-Biggs2000}%
  \BibitemOpen
  \bibfield  {author} {\bibinfo {author} {\bibfnamefont {Michael}\ \bibnamefont
  {Bartholomew-Biggs}}, \bibinfo {author} {\bibfnamefont {Steven}\ \bibnamefont
  {Brown}}, \bibinfo {author} {\bibfnamefont {Bruce}\ \bibnamefont
  {Christianson}}, \ and\ \bibinfo {author} {\bibfnamefont {Laurence}\
  \bibnamefont {Dixon}},\ }\bibfield  {title} {\enquote {\bibinfo {title}
  {{Automatic differentiation of algorithms}},}\ }\href {\doibase
  10.1016/S0377-0427(00)00422-2} {\bibfield  {journal} {\bibinfo  {journal} {J.
  Comput. Appl. Math.}\ }\textbf {\bibinfo {volume} {124}},\ \bibinfo {pages}
  {171--190} (\bibinfo {year} {2000})}\BibitemShut {NoStop}%
\bibitem [{\citenamefont {Baur}\ and\ \citenamefont
  {Strassen}(1983)}]{Baur1983}%
  \BibitemOpen
  \bibfield  {author} {\bibinfo {author} {\bibfnamefont {Walter}\ \bibnamefont
  {Baur}}\ and\ \bibinfo {author} {\bibfnamefont {Volker}\ \bibnamefont
  {Strassen}},\ }\bibfield  {title} {\enquote {\bibinfo {title} {{The
  Complexity of Partial Derivatives}},}\ }\href {\doibase
  https://doi.org/10.1016/0304-3975(83)90110-X} {\bibfield  {journal} {\bibinfo
   {journal} {Theor. Comput. Sci.}\ }\textbf {\bibinfo {volume} {22}},\
  \bibinfo {pages} {317} (\bibinfo {year} {1983})}\BibitemShut {NoStop}%
\bibitem [{\citenamefont {Griewank}(1989)}]{Griewank1989}%
  \BibitemOpen
  \bibfield  {author} {\bibinfo {author} {\bibfnamefont {Andreas}\ \bibnamefont
  {Griewank}},\ }\bibfield  {title} {\enquote {\bibinfo {title} {{On Automatic
  Differentiation}},}\ }in\ \href
  {http://citeseerx.ist.psu.edu/viewdoc/summary?doi=10.1.1.54.2317} {\emph
  {\bibinfo {booktitle} {Math. Program. Recent Dev. Appl.}}}\ (\bibinfo
  {publisher} {Kluwer Academic Publishers},\ \bibinfo {year} {1989})\ pp.\
  \bibinfo {pages} {83--108}\BibitemShut {NoStop}%
\bibitem [{\citenamefont {Rumelhart}\ \emph {et~al.}(1986)\citenamefont
  {Rumelhart}, \citenamefont {Hinton},\ and\ \citenamefont
  {Williams}}]{BP1986}%
  \BibitemOpen
  \bibfield  {author} {\bibinfo {author} {\bibfnamefont {David~E.}\
  \bibnamefont {Rumelhart}}, \bibinfo {author} {\bibfnamefont {Geoffrey~E.}\
  \bibnamefont {Hinton}}, \ and\ \bibinfo {author} {\bibfnamefont {Ronald~J.}\
  \bibnamefont {Williams}},\ }\bibfield  {title} {\enquote {\bibinfo {title}
  {{Learning representations by back-propagating errors}},}\ }\href
  {https://www.nature.com/articles/323533a0} {\bibfield  {journal} {\bibinfo
  {journal} {Nature}\ }\textbf {\bibinfo {volume} {323}},\ \bibinfo {pages}
  {533} (\bibinfo {year} {1986})}\BibitemShut {NoStop}%
\bibitem [{\citenamefont {Baydin}\ \emph {et~al.}(2015)\citenamefont {Baydin},
  \citenamefont {Pearlmutter}, \citenamefont {Radul},\ and\ \citenamefont
  {Siskind}}]{Baydin2018}%
  \BibitemOpen
  \bibfield  {author} {\bibinfo {author} {\bibfnamefont {Atilim~G\"unes}\
  \bibnamefont {Baydin}}, \bibinfo {author} {\bibfnamefont {Barak~A.}\
  \bibnamefont {Pearlmutter}}, \bibinfo {author} {\bibfnamefont
  {Alexey~Andreyevich}\ \bibnamefont {Radul}}, \ and\ \bibinfo {author}
  {\bibfnamefont {Jeffrey~Mark}\ \bibnamefont {Siskind}},\ }\bibfield  {title}
  {\enquote {\bibinfo {title} {{Automatic differentiation in machine learning:
  a survey}},}\ }\href {\doibase 10.1016/j.advwatres.2018.01.009} {\bibfield
  {journal} {\bibinfo  {journal} {J. Mach. Learn.}\ }\textbf {\bibinfo {volume}
  {18}},\ \bibinfo {pages} {1--43} (\bibinfo {year} {2015})},\ \Eprint
  {http://arxiv.org/abs/1502.05767} {arXiv:1502.05767} \BibitemShut {NoStop}%
\bibitem [{\citenamefont {Leung}\ \emph {et~al.}(2017)\citenamefont {Leung},
  \citenamefont {Abdelhafez}, \citenamefont {Koch},\ and\ \citenamefont
  {Schuster}}]{Leung2017}%
  \BibitemOpen
  \bibfield  {author} {\bibinfo {author} {\bibfnamefont {Nelson}\ \bibnamefont
  {Leung}}, \bibinfo {author} {\bibfnamefont {Mohamed}\ \bibnamefont
  {Abdelhafez}}, \bibinfo {author} {\bibfnamefont {Jens}\ \bibnamefont {Koch}},
  \ and\ \bibinfo {author} {\bibfnamefont {David}\ \bibnamefont {Schuster}},\
  }\bibfield  {title} {\enquote {\bibinfo {title} {{Speedup for quantum optimal
  control from automatic differentiation based on graphics processing
  units}},}\ }\href {\doibase 10.1103/PhysRevA.95.042318} {\bibfield  {journal}
  {\bibinfo  {journal} {Phys. Rev. A}\ }\textbf {\bibinfo {volume} {95}},\
  \bibinfo {pages} {042318} (\bibinfo {year} {2017})},\ \Eprint
  {http://arxiv.org/abs/1612.04929} {arXiv:1612.04929} \BibitemShut {NoStop}%
\bibitem [{\citenamefont {Sorella}\ and\ \citenamefont
  {Capriotti}(2010)}]{Sorella2010}%
  \BibitemOpen
  \bibfield  {author} {\bibinfo {author} {\bibfnamefont {Sandro}\ \bibnamefont
  {Sorella}}\ and\ \bibinfo {author} {\bibfnamefont {Luca}\ \bibnamefont
  {Capriotti}},\ }\bibfield  {title} {\enquote {\bibinfo {title} {{Algorithmic
  differentiation and the calculation of forces by quantum Monte Carlo}},}\
  }\href@noop {} {\bibfield  {journal} {\bibinfo  {journal} {J. Chem. Phys.}\
  }\textbf {\bibinfo {volume} {133}} (\bibinfo {year} {2010})}\BibitemShut
  {NoStop}%
\bibitem [{\citenamefont {Tamayo-Mendoza}\ \emph {et~al.}(2018)\citenamefont
  {Tamayo-Mendoza}, \citenamefont {Kreisbeck}, \citenamefont {Lindh},\ and\
  \citenamefont {Aspuru-Guzik}}]{Tamayo-Mendoza2018}%
  \BibitemOpen
  \bibfield  {author} {\bibinfo {author} {\bibfnamefont {Teresa}\ \bibnamefont
  {Tamayo-Mendoza}}, \bibinfo {author} {\bibfnamefont {Christoph}\ \bibnamefont
  {Kreisbeck}}, \bibinfo {author} {\bibfnamefont {Roland}\ \bibnamefont
  {Lindh}}, \ and\ \bibinfo {author} {\bibfnamefont {Al\'an}\ \bibnamefont
  {Aspuru-Guzik}},\ }\bibfield  {title} {\enquote {\bibinfo {title} {{Automatic
  Differentiation in Quantum Chemistry with Applications to Fully Variational
  Hartree-Fock}},}\ }\href {\doibase 10.1021/acscentsci.7b00586} {\bibfield
  {journal} {\bibinfo  {journal} {ACS Cent. Sci.}\ }\textbf {\bibinfo {volume}
  {4}},\ \bibinfo {pages} {559--566} (\bibinfo {year} {2018})}\BibitemShut
  {NoStop}%
\bibitem [{Note1()}]{Note1}%
  \BibitemOpen
  \bibinfo {note} {As noted in \cite {NIPS2018_7540} one will need tensor
  calculus as a precise language to present automatic differentiation, even the
  object being differentiated is a matrix.}\BibitemShut {Stop}%
\bibitem [{\citenamefont {Griewank}\ and\ \citenamefont
  {Walther}(2008)}]{Griewank2008}%
  \BibitemOpen
  \bibfield  {author} {\bibinfo {author} {\bibfnamefont {Andreas}\ \bibnamefont
  {Griewank}}\ and\ \bibinfo {author} {\bibfnamefont {Andrea}\ \bibnamefont
  {Walther}},\ }\href {\doibase 10.1137/1.9780898717761} {\emph {\bibinfo
  {title} {{Evaluating Derivatives}}}}\ (\bibinfo  {publisher} {Society for
  Industrial and Applied Mathematics},\ \bibinfo {year} {2008})\BibitemShut
  {NoStop}%
\bibitem [{\citenamefont {Abadi}\ \emph {et~al.}(2016)\citenamefont {Abadi},
  \citenamefont {Agarwal}, \citenamefont {Barham}, \citenamefont {Brevdo},
  \citenamefont {Chen}, \citenamefont {Citro}, \citenamefont {Corrado},
  \citenamefont {Davis}, \citenamefont {Dean}, \citenamefont {Devin},
  \citenamefont {Ghemawat}, \citenamefont {Goodfellow}, \citenamefont {Harp},
  \citenamefont {Irving}, \citenamefont {Isard}, \citenamefont {Jia},
  \citenamefont {Jozefowicz}, \citenamefont {Kaiser}, \citenamefont {Kudlur},
  \citenamefont {Levenberg}, \citenamefont {Mane}, \citenamefont {Monga},
  \citenamefont {Moore}, \citenamefont {Murray}, \citenamefont {Olah},
  \citenamefont {Schuster}, \citenamefont {Shlens}, \citenamefont {Steiner},
  \citenamefont {Sutskever}, \citenamefont {Talwar}, \citenamefont {Tucker},
  \citenamefont {Vanhoucke}, \citenamefont {Vasudevan}, \citenamefont {Viegas},
  \citenamefont {Vinyals}, \citenamefont {Warden}, \citenamefont {Wattenberg},
  \citenamefont {Wicke}, \citenamefont {Yu},\ and\ \citenamefont
  {Zheng}}]{Abadi2016}%
  \BibitemOpen
  \bibfield  {author} {\bibinfo {author} {\bibfnamefont {Martin}\ \bibnamefont
  {Abadi}}, \bibinfo {author} {\bibfnamefont {Ashish}\ \bibnamefont {Agarwal}},
  \bibinfo {author} {\bibfnamefont {Paul}\ \bibnamefont {Barham}}, \bibinfo
  {author} {\bibfnamefont {Eugene}\ \bibnamefont {Brevdo}}, \bibinfo {author}
  {\bibfnamefont {Zhifeng}\ \bibnamefont {Chen}}, \bibinfo {author}
  {\bibfnamefont {Craig}\ \bibnamefont {Citro}}, \bibinfo {author}
  {\bibfnamefont {Greg~S.}\ \bibnamefont {Corrado}}, \bibinfo {author}
  {\bibfnamefont {Andy}\ \bibnamefont {Davis}}, \bibinfo {author}
  {\bibfnamefont {Jeffrey}\ \bibnamefont {Dean}}, \bibinfo {author}
  {\bibfnamefont {Matthieu}\ \bibnamefont {Devin}}, \bibinfo {author}
  {\bibfnamefont {Sanjay}\ \bibnamefont {Ghemawat}}, \bibinfo {author}
  {\bibfnamefont {Ian}\ \bibnamefont {Goodfellow}}, \bibinfo {author}
  {\bibfnamefont {Andrew}\ \bibnamefont {Harp}}, \bibinfo {author}
  {\bibfnamefont {Geoffrey}\ \bibnamefont {Irving}}, \bibinfo {author}
  {\bibfnamefont {Michael}\ \bibnamefont {Isard}}, \bibinfo {author}
  {\bibfnamefont {Yangqing}\ \bibnamefont {Jia}}, \bibinfo {author}
  {\bibfnamefont {Rafal}\ \bibnamefont {Jozefowicz}}, \bibinfo {author}
  {\bibfnamefont {Lukasz}\ \bibnamefont {Kaiser}}, \bibinfo {author}
  {\bibfnamefont {Manjunath}\ \bibnamefont {Kudlur}}, \bibinfo {author}
  {\bibfnamefont {Josh}\ \bibnamefont {Levenberg}}, \bibinfo {author}
  {\bibfnamefont {Dan}\ \bibnamefont {Mane}}, \bibinfo {author} {\bibfnamefont
  {Rajat}\ \bibnamefont {Monga}}, \bibinfo {author} {\bibfnamefont {Sherry}\
  \bibnamefont {Moore}}, \bibinfo {author} {\bibfnamefont {Derek}\ \bibnamefont
  {Murray}}, \bibinfo {author} {\bibfnamefont {Chris}\ \bibnamefont {Olah}},
  \bibinfo {author} {\bibfnamefont {Mike}\ \bibnamefont {Schuster}}, \bibinfo
  {author} {\bibfnamefont {Jonathon}\ \bibnamefont {Shlens}}, \bibinfo {author}
  {\bibfnamefont {Benoit}\ \bibnamefont {Steiner}}, \bibinfo {author}
  {\bibfnamefont {Ilya}\ \bibnamefont {Sutskever}}, \bibinfo {author}
  {\bibfnamefont {Kunal}\ \bibnamefont {Talwar}}, \bibinfo {author}
  {\bibfnamefont {Paul}\ \bibnamefont {Tucker}}, \bibinfo {author}
  {\bibfnamefont {Vincent}\ \bibnamefont {Vanhoucke}}, \bibinfo {author}
  {\bibfnamefont {Vijay}\ \bibnamefont {Vasudevan}}, \bibinfo {author}
  {\bibfnamefont {Fernanda}\ \bibnamefont {Viegas}}, \bibinfo {author}
  {\bibfnamefont {Oriol}\ \bibnamefont {Vinyals}}, \bibinfo {author}
  {\bibfnamefont {Pete}\ \bibnamefont {Warden}}, \bibinfo {author}
  {\bibfnamefont {Martin}\ \bibnamefont {Wattenberg}}, \bibinfo {author}
  {\bibfnamefont {Martin}\ \bibnamefont {Wicke}}, \bibinfo {author}
  {\bibfnamefont {Yuan}\ \bibnamefont {Yu}}, \ and\ \bibinfo {author}
  {\bibfnamefont {Xiaoqiang}\ \bibnamefont {Zheng}},\ }\bibfield  {title}
  {\enquote {\bibinfo {title} {{TensorFlow: Large-Scale Machine Learning on
  Heterogeneous Distributed Systems}},}\ }\href
  {http://arxiv.org/abs/1603.04467} {\bibfield  {journal} {\bibinfo  {journal}
  {arXiv}\ } (\bibinfo {year} {2016})},\ \Eprint
  {http://arxiv.org/abs/1603.04467} {arXiv:1603.04467} \BibitemShut {NoStop}%
\bibitem [{\citenamefont {Maclaurin}\ \emph {et~al.}(2015)\citenamefont
  {Maclaurin}, \citenamefont {Duvenaud},\ and\ \citenamefont
  {Adams}}]{maclaurin2015autograd}%
  \BibitemOpen
  \bibfield  {author} {\bibinfo {author} {\bibfnamefont {Dougal}\ \bibnamefont
  {Maclaurin}}, \bibinfo {author} {\bibfnamefont {David}\ \bibnamefont
  {Duvenaud}}, \ and\ \bibinfo {author} {\bibfnamefont {Ryan~P}\ \bibnamefont
  {Adams}},\ }\bibfield  {title} {\enquote {\bibinfo {title} {Autograd:
  Effortless gradients in numpy},}\ }in\ \href@noop {} {\emph {\bibinfo
  {booktitle} {ICML 2015 AutoML Workshop}}}\ (\bibinfo {year}
  {2015})\BibitemShut {NoStop}%
\bibitem [{\citenamefont {Paszke}\ \emph {et~al.}(2017)\citenamefont {Paszke},
  \citenamefont {Chanan}, \citenamefont {Lin}, \citenamefont {Gross},
  \citenamefont {Yang}, \citenamefont {Antiga},\ and\ \citenamefont
  {Devito}}]{Paszke2017}%
  \BibitemOpen
  \bibfield  {author} {\bibinfo {author} {\bibfnamefont {Adam}\ \bibnamefont
  {Paszke}}, \bibinfo {author} {\bibfnamefont {Gregory}\ \bibnamefont
  {Chanan}}, \bibinfo {author} {\bibfnamefont {Zeming}\ \bibnamefont {Lin}},
  \bibinfo {author} {\bibfnamefont {Sam}\ \bibnamefont {Gross}}, \bibinfo
  {author} {\bibfnamefont {Edward}\ \bibnamefont {Yang}}, \bibinfo {author}
  {\bibfnamefont {Luca}\ \bibnamefont {Antiga}}, \ and\ \bibinfo {author}
  {\bibfnamefont {Zachary}\ \bibnamefont {Devito}},\ }\bibfield  {title}
  {\enquote {\bibinfo {title} {{Automatic differentiation in PyTorch}},}\
  }\href@noop {} {\bibfield  {journal} {\bibinfo  {journal} {31st Conf. Neural
  Inf. Process. Syst.}\ } (\bibinfo {year} {2017})},\ \Eprint
  {http://arxiv.org/abs/1011.1669} {arXiv:1011.1669} \BibitemShut {NoStop}%
\bibitem [{\citenamefont {Johnson}\ \emph {et~al.}(2018)\citenamefont
  {Johnson}, \citenamefont {Frostig},\ and\ \citenamefont {Leary}}]{jaxpaper}%
  \BibitemOpen
  \bibfield  {author} {\bibinfo {author} {\bibfnamefont {Matthew}\ \bibnamefont
  {Johnson}}, \bibinfo {author} {\bibfnamefont {Roy}\ \bibnamefont {Frostig}},
  \ and\ \bibinfo {author} {\bibfnamefont {Chris}\ \bibnamefont {Leary}},\
  }\bibfield  {title} {\enquote {\bibinfo {title} {Compiling machine learning
  programs via high-level tracing},}\ }in\ \href
  {http://www.sysml.cc/doc/146.pdf} {\emph {\bibinfo {booktitle} {SysML}}}\
  (\bibinfo {year} {2018})\BibitemShut {NoStop}%
\bibitem [{\citenamefont {Innes}(2018)}]{innes2018don}%
  \BibitemOpen
  \bibfield  {author} {\bibinfo {author} {\bibfnamefont {Michael}\ \bibnamefont
  {Innes}},\ }\bibfield  {title} {\enquote {\bibinfo {title} {Don't unroll
  adjoint: Differentiating ssa-form programs},}\ }\href@noop {} {\bibfield
  {journal} {\bibinfo  {journal} {arXiv}\ } (\bibinfo {year} {2018})},\ \Eprint
  {http://arxiv.org/abs/1810.07951} {arXiv:1810.07951} \BibitemShut {NoStop}%
\bibitem [{\citenamefont {Kourtis}\ \emph {et~al.}(2018)\citenamefont
  {Kourtis}, \citenamefont {Chamon}, \citenamefont {Mucciolo},\ and\
  \citenamefont {Ruckenstein}}]{Kourtis2018}%
  \BibitemOpen
  \bibfield  {author} {\bibinfo {author} {\bibfnamefont {Stefanos}\
  \bibnamefont {Kourtis}}, \bibinfo {author} {\bibfnamefont {Claudio}\
  \bibnamefont {Chamon}}, \bibinfo {author} {\bibfnamefont {Eduardo~R}\
  \bibnamefont {Mucciolo}}, \ and\ \bibinfo {author} {\bibfnamefont {Andrei~E}\
  \bibnamefont {Ruckenstein}},\ }\bibfield  {title} {\enquote {\bibinfo {title}
  {{Fast counting with tensor networks}},}\ }\href
  {http://arxiv.org/abs/1805.00475} {\bibfield  {journal} {\bibinfo  {journal}
  {arXiv}\ } (\bibinfo {year} {2018})},\ \Eprint
  {http://arxiv.org/abs/1805.00475} {arXiv:1805.00475} \BibitemShut {NoStop}%
\bibitem [{\citenamefont {Villalonga}\ \emph {et~al.}(2018)\citenamefont
  {Villalonga}, \citenamefont {Boixo}, \citenamefont {Nelson}, \citenamefont
  {Rieffel}, \citenamefont {Biswas},\ and\ \citenamefont
  {Mandr}}]{Villalonga2018}%
  \BibitemOpen
  \bibfield  {author} {\bibinfo {author} {\bibfnamefont {Benjamin}\
  \bibnamefont {Villalonga}}, \bibinfo {author} {\bibfnamefont {Sergio}\
  \bibnamefont {Boixo}}, \bibinfo {author} {\bibfnamefont {Bron}\ \bibnamefont
  {Nelson}}, \bibinfo {author} {\bibfnamefont {Eleanor}\ \bibnamefont
  {Rieffel}}, \bibinfo {author} {\bibfnamefont {Rupak}\ \bibnamefont {Biswas}},
  \ and\ \bibinfo {author} {\bibfnamefont {Salvatore}\ \bibnamefont {Mandr}},\
  }\bibfield  {title} {\enquote {\bibinfo {title} {{A flexible high-performance
  simulator for the verification and benchmarking of quantum circuits
  implemented on real hardware Benjamin}},}\ }\href
  {https://arxiv.org/abs/1811.09599} {\bibfield  {journal} {\bibinfo  {journal}
  {arXiv}\ } (\bibinfo {year} {2018})},\ \Eprint
  {http://arxiv.org/abs/1811.09599v1} {arXiv:1811.09599v1} \BibitemShut
  {NoStop}%
\bibitem [{\citenamefont {Levin}\ and\ \citenamefont {Nave}(2007)}]{Levin2007}%
  \BibitemOpen
  \bibfield  {author} {\bibinfo {author} {\bibfnamefont {Michael}\ \bibnamefont
  {Levin}}\ and\ \bibinfo {author} {\bibfnamefont {Cody~P}\ \bibnamefont
  {Nave}},\ }\bibfield  {title} {\enquote {\bibinfo {title} {{Tensor
  renormalization group approach to two-dimensional classical lattice
  models}},}\ }\href {\doibase 10.1103/PhysRevLett.99.120601} {\bibfield
  {journal} {\bibinfo  {journal} {Phys. Rev. Lett.}\ }\textbf {\bibinfo
  {volume} {99}},\ \bibinfo {pages} {120601} (\bibinfo {year}
  {2007})}\BibitemShut {NoStop}%
\bibitem [{\citenamefont {Xie}\ \emph {et~al.}(2008)\citenamefont {Xie},
  \citenamefont {Jiang}, \citenamefont {Chen}, \citenamefont {Weng},\ and\
  \citenamefont {Xiang}}]{Xie2008}%
  \BibitemOpen
  \bibfield  {author} {\bibinfo {author} {\bibfnamefont {Z.~Y.}\ \bibnamefont
  {Xie}}, \bibinfo {author} {\bibfnamefont {H.~C.}\ \bibnamefont {Jiang}},
  \bibinfo {author} {\bibfnamefont {Q.~N.}\ \bibnamefont {Chen}}, \bibinfo
  {author} {\bibfnamefont {Z.~Y.}\ \bibnamefont {Weng}}, \ and\ \bibinfo
  {author} {\bibfnamefont {Tao}\ \bibnamefont {Xiang}},\ }\bibfield  {title}
  {\enquote {\bibinfo {title} {{Second Renormalization of Tensor-Network
  States}},}\ }\href {\doibase 10.1103/PhysRevLett.103.160601} {\bibfield
  {journal} {\bibinfo  {journal} {Phys. Rev. Lett.}\ }\textbf {\bibinfo
  {volume} {103}},\ \bibinfo {pages} {160601} (\bibinfo {year} {2008})},\
  \Eprint {http://arxiv.org/abs/0809.0182} {arXiv:0809.0182} \BibitemShut
  {NoStop}%
\bibitem [{\citenamefont {Gu}\ and\ \citenamefont {Wen}(2009)}]{Gu2009}%
  \BibitemOpen
  \bibfield  {author} {\bibinfo {author} {\bibfnamefont {Zheng-Cheng}\
  \bibnamefont {Gu}}\ and\ \bibinfo {author} {\bibfnamefont {Xiao-Gang}\
  \bibnamefont {Wen}},\ }\bibfield  {title} {\enquote {\bibinfo {title}
  {{Tensor-entanglement-filtering renormalization approach and
  symmetry-protected topological order}},}\ }\href {\doibase
  10.1103/PhysRevB.80.155131} {\bibfield  {journal} {\bibinfo  {journal} {Phys.
  Rev. B}\ }\textbf {\bibinfo {volume} {80}},\ \bibinfo {pages} {155131}
  (\bibinfo {year} {2009})},\ \Eprint {http://arxiv.org/abs/0903.1069}
  {arXiv:0903.1069} \BibitemShut {NoStop}%
\bibitem [{\citenamefont {Xie}\ \emph {et~al.}(2012)\citenamefont {Xie},
  \citenamefont {Chen}, \citenamefont {Qin}, \citenamefont {Zhu}, \citenamefont
  {Yang},\ and\ \citenamefont {Xiang}}]{Xie2012}%
  \BibitemOpen
  \bibfield  {author} {\bibinfo {author} {\bibfnamefont {Z.~Y.}\ \bibnamefont
  {Xie}}, \bibinfo {author} {\bibfnamefont {Jing}\ \bibnamefont {Chen}},
  \bibinfo {author} {\bibfnamefont {M.~P.}\ \bibnamefont {Qin}}, \bibinfo
  {author} {\bibfnamefont {J.~W.}\ \bibnamefont {Zhu}}, \bibinfo {author}
  {\bibfnamefont {L.~P.}\ \bibnamefont {Yang}}, \ and\ \bibinfo {author}
  {\bibfnamefont {Tao}\ \bibnamefont {Xiang}},\ }\bibfield  {title} {\enquote
  {\bibinfo {title} {{Coarse-graining renormalization by higher-order singular
  value decomposition}},}\ }\href {\doibase 10.1103/PhysRevB.86.045139}
  {\bibfield  {journal} {\bibinfo  {journal} {Phys. Rev. B}\ }\textbf {\bibinfo
  {volume} {86}},\ \bibinfo {pages} {045139} (\bibinfo {year} {2012})},\
  \Eprint {http://arxiv.org/abs/1201.1144} {arXiv:1201.1144} \BibitemShut
  {NoStop}%
\bibitem [{\citenamefont {Xie}\ \emph {et~al.}(2014)\citenamefont {Xie},
  \citenamefont {Chen}, \citenamefont {Yu}, \citenamefont {Kong}, \citenamefont
  {Normand},\ and\ \citenamefont {Xiang}}]{Xie2014}%
  \BibitemOpen
  \bibfield  {author} {\bibinfo {author} {\bibfnamefont {Z.~Y.}\ \bibnamefont
  {Xie}}, \bibinfo {author} {\bibfnamefont {J.}~\bibnamefont {Chen}}, \bibinfo
  {author} {\bibfnamefont {J.~F.}\ \bibnamefont {Yu}}, \bibinfo {author}
  {\bibfnamefont {X.}~\bibnamefont {Kong}}, \bibinfo {author} {\bibfnamefont
  {B.}~\bibnamefont {Normand}}, \ and\ \bibinfo {author} {\bibfnamefont
  {T.}~\bibnamefont {Xiang}},\ }\bibfield  {title} {\enquote {\bibinfo {title}
  {{Tensor Renormalization of Quantum Many-Body Systems Using Projected
  Entangled Simplex States}},}\ }\href {\doibase 10.1103/physrevx.4.011025}
  {\bibfield  {journal} {\bibinfo  {journal} {Phys. Rev. X}\ }\textbf {\bibinfo
  {volume} {4}},\ \bibinfo {pages} {011025} (\bibinfo {year}
  {2014})}\BibitemShut {NoStop}%
\bibitem [{\citenamefont {Evenbly}\ and\ \citenamefont
  {Vidal}(2015)}]{Evenbly2015}%
  \BibitemOpen
  \bibfield  {author} {\bibinfo {author} {\bibfnamefont {G.}~\bibnamefont
  {Evenbly}}\ and\ \bibinfo {author} {\bibfnamefont {G.}~\bibnamefont
  {Vidal}},\ }\bibfield  {title} {\enquote {\bibinfo {title} {{Tensor Network
  Renormalization}},}\ }\href {\doibase 10.1103/PhysRevLett.115.180405}
  {\bibfield  {journal} {\bibinfo  {journal} {Phys. Rev. Lett.}\ }\textbf
  {\bibinfo {volume} {115}},\ \bibinfo {pages} {180405} (\bibinfo {year}
  {2015})}\BibitemShut {NoStop}%
\bibitem [{\citenamefont {Yang}\ \emph {et~al.}(2017)\citenamefont {Yang},
  \citenamefont {Gu},\ and\ \citenamefont {Wen}}]{Yang2017c}%
  \BibitemOpen
  \bibfield  {author} {\bibinfo {author} {\bibfnamefont {Shuo}\ \bibnamefont
  {Yang}}, \bibinfo {author} {\bibfnamefont {Zheng-Cheng}\ \bibnamefont {Gu}},
  \ and\ \bibinfo {author} {\bibfnamefont {Xiao-Gang}\ \bibnamefont {Wen}},\
  }\bibfield  {title} {\enquote {\bibinfo {title} {{Loop Optimization for
  Tensor Network Renormalization}},}\ }\href {\doibase
  10.1103/PhysRevLett.118.110504} {\bibfield  {journal} {\bibinfo  {journal}
  {Phys. Rev. Lett.}\ }\textbf {\bibinfo {volume} {118}},\ \bibinfo {pages}
  {110504} (\bibinfo {year} {2017})}\BibitemShut {NoStop}%
\bibitem [{\citenamefont {Nishino}\ and\ \citenamefont
  {Okunishi}(1996)}]{Nishino1996}%
  \BibitemOpen
  \bibfield  {author} {\bibinfo {author} {\bibfnamefont {Tomotoshi}\
  \bibnamefont {Nishino}}\ and\ \bibinfo {author} {\bibfnamefont {Kouichi}\
  \bibnamefont {Okunishi}},\ }\bibfield  {title} {\enquote {\bibinfo {title}
  {{Corner Transfer Matrix Renormalization Group Method}},}\ }\href@noop {}
  {\bibfield  {journal} {\bibinfo  {journal} {J. Phys. Soc. Jpn.}\ }\textbf
  {\bibinfo {volume} {65}},\ \bibinfo {pages} {891} (\bibinfo {year}
  {1996})}\BibitemShut {NoStop}%
\bibitem [{\citenamefont {Or{\'{u}}s}(2012)}]{Orus2012}%
  \BibitemOpen
  \bibfield  {author} {\bibinfo {author} {\bibfnamefont {Rom{\'{a}}n}\
  \bibnamefont {Or{\'{u}}s}},\ }\bibfield  {title} {\enquote {\bibinfo {title}
  {{Exploring corner transfer matrices and corner tensors for the classical
  simulation of quantum lattice systems}},}\ }\href {\doibase
  10.1103/PhysRevB.85.205117} {\bibfield  {journal} {\bibinfo  {journal} {Phys.
  Rev. B}\ }\textbf {\bibinfo {volume} {85}},\ \bibinfo {pages} {205117}
  (\bibinfo {year} {2012})},\ \Eprint {http://arxiv.org/abs/1112.4101}
  {arXiv:1112.4101} \BibitemShut {NoStop}%
\bibitem [{\citenamefont {Corboz}\ \emph {et~al.}(2014)\citenamefont {Corboz},
  \citenamefont {Rice},\ and\ \citenamefont {Troyer}}]{Corboz2014}%
  \BibitemOpen
  \bibfield  {author} {\bibinfo {author} {\bibfnamefont {Philippe}\
  \bibnamefont {Corboz}}, \bibinfo {author} {\bibfnamefont {T.~M.}\
  \bibnamefont {Rice}}, \ and\ \bibinfo {author} {\bibfnamefont {Matthias}\
  \bibnamefont {Troyer}},\ }\bibfield  {title} {\enquote {\bibinfo {title}
  {{Competing States in the t-J Model: Uniform d-Wave State versus Stripe State
  Philippe}},}\ }\href {\doibase 10.1103/PhysRevLett.113.046402} {\bibfield
  {journal} {\bibinfo  {journal} {Phys. Rev. Lett.}\ }\textbf {\bibinfo
  {volume} {113}},\ \bibinfo {pages} {046402} (\bibinfo {year}
  {2014})}\BibitemShut {NoStop}%
\bibitem [{\citenamefont {Orus}\ and\ \citenamefont {Vidal}(2008)}]{Orus2008}%
  \BibitemOpen
  \bibfield  {author} {\bibinfo {author} {\bibfnamefont {R.}~\bibnamefont
  {Orus}}\ and\ \bibinfo {author} {\bibfnamefont {G.}~\bibnamefont {Vidal}},\
  }\bibfield  {title} {\enquote {\bibinfo {title} {{Infinite time-evolving
  block decimation algorithm beyond unitary evolution}},}\ }\href {\doibase
  10.1103/PhysRevB.78.155117} {\bibfield  {journal} {\bibinfo  {journal} {Phys.
  Rev. B}\ }\textbf {\bibinfo {volume} {78}},\ \bibinfo {pages} {155117}
  (\bibinfo {year} {2008})}\BibitemShut {NoStop}%
\bibitem [{\citenamefont {Giles}(2008)}]{Giles2008}%
  \BibitemOpen
  \bibfield  {author} {\bibinfo {author} {\bibfnamefont {Mike}\ \bibnamefont
  {Giles}},\ }\href {https://people.maths.ox.ac.uk/gilesm/files/NA-08-01.pdf}
  {\emph {\bibinfo {title} {{An extended collection of matrix derivative
  results for forward and reverse mode algorithmic differentiation}}}},\
  \bibinfo {type} {Tech. Rep.}\ (\bibinfo {year} {2008})\BibitemShut {NoStop}%
\bibitem [{\citenamefont {Townsend}(2016)}]{Townsend}%
  \BibitemOpen
  \bibfield  {author} {\bibinfo {author} {\bibfnamefont {James}\ \bibnamefont
  {Townsend}},\ }\href {https://j-towns.github.io/papers/svd-derivative.pdf}
  {\emph {\bibinfo {title} {{Differentiating the Singular Value
  Decomposition}}}},\ \bibinfo {type} {Tech. Rep.}\ (\bibinfo {year}
  {2016})\BibitemShut {NoStop}%
\bibitem [{\citenamefont {Seeger}\ \emph {et~al.}(2017)\citenamefont {Seeger},
  \citenamefont {Hetzel}, \citenamefont {Dai}, \citenamefont {Meissner},\ and\
  \citenamefont {Lawrence}}]{Seeger2017}%
  \BibitemOpen
  \bibfield  {author} {\bibinfo {author} {\bibfnamefont {Matthias}\
  \bibnamefont {Seeger}}, \bibinfo {author} {\bibfnamefont {Asmus}\
  \bibnamefont {Hetzel}}, \bibinfo {author} {\bibfnamefont {Zhenwen}\
  \bibnamefont {Dai}}, \bibinfo {author} {\bibfnamefont {Eric}\ \bibnamefont
  {Meissner}}, \ and\ \bibinfo {author} {\bibfnamefont {Neil~D.}\ \bibnamefont
  {Lawrence}},\ }\bibfield  {title} {\enquote {\bibinfo {title}
  {{Auto-Differentiating Linear Algebra}},}\ }\href
  {http://arxiv.org/abs/1710.08717} {\bibfield  {journal} {\bibinfo  {journal}
  {arXiv}\ } (\bibinfo {year} {2017})},\ \Eprint
  {http://arxiv.org/abs/1710.08717} {arXiv:1710.08717} \BibitemShut {NoStop}%
\bibitem [{\citenamefont {Feynman}(1939)}]{Feynman1939}%
  \BibitemOpen
  \bibfield  {author} {\bibinfo {author} {\bibfnamefont {Richard~P.}\
  \bibnamefont {Feynman}},\ }\bibfield  {title} {\enquote {\bibinfo {title}
  {{Forces on Molecules}},}\ }\href {\doibase 10.1103/PhysRev.56.340}
  {\bibfield  {journal} {\bibinfo  {journal} {Phys. Rev.}\ }\textbf {\bibinfo
  {volume} {56}},\ \bibinfo {pages} {340--343} (\bibinfo {year}
  {1939})}\BibitemShut {NoStop}%
\bibitem [{\citenamefont {Fujita}\ \emph {et~al.}(2018)\citenamefont {Fujita},
  \citenamefont {Nakagawa}, \citenamefont {Sugiura},\ and\ \citenamefont
  {Oshikawa}}]{Fujita2018}%
  \BibitemOpen
  \bibfield  {author} {\bibinfo {author} {\bibfnamefont {Hiroyuki}\
  \bibnamefont {Fujita}}, \bibinfo {author} {\bibfnamefont {Yuya~O}\
  \bibnamefont {Nakagawa}}, \bibinfo {author} {\bibfnamefont {Sho}\
  \bibnamefont {Sugiura}}, \ and\ \bibinfo {author} {\bibfnamefont {Masaki}\
  \bibnamefont {Oshikawa}},\ }\bibfield  {title} {\enquote {\bibinfo {title}
  {{Construction of Hamiltonians by supervised learning of energy and
  entanglement spectra}},}\ }\href {\doibase 10.1103/PhysRevB.97.075114}
  {\bibfield  {journal} {\bibinfo  {journal} {Phys. Rev. B}\ }\textbf {\bibinfo
  {volume} {97}},\ \bibinfo {pages} {075114} (\bibinfo {year}
  {2018})}\BibitemShut {NoStop}%
\bibitem [{\citenamefont {Martin}(2004)}]{martin2004electronic}%
  \BibitemOpen
  \bibfield  {author} {\bibinfo {author} {\bibfnamefont {Richard~M}\
  \bibnamefont {Martin}},\ }\href@noop {} {\emph {\bibinfo {title} {Electronic
  structure: basic theory and practical methods}}}\ (\bibinfo  {publisher}
  {Cambridge university press},\ \bibinfo {year} {2004})\BibitemShut {NoStop}%
\bibitem [{Note2()}]{Note2}%
  \BibitemOpen
  \bibinfo {note} {The cases where one has exact degeneracy between occupied
  and unoccupied eigenstates is itself a singular case. One should include or
  exclude all degenerated states in these cases.}\BibitemShut {Stop}%
\bibitem [{\citenamefont {Pollmann}\ \emph {et~al.}(2010)\citenamefont
  {Pollmann}, \citenamefont {Turner}, \citenamefont {Berg},\ and\ \citenamefont
  {Oshikawa}}]{Pollmann2010}%
  \BibitemOpen
  \bibfield  {author} {\bibinfo {author} {\bibfnamefont {Frank}\ \bibnamefont
  {Pollmann}}, \bibinfo {author} {\bibfnamefont {Ari~M.}\ \bibnamefont
  {Turner}}, \bibinfo {author} {\bibfnamefont {Erez}\ \bibnamefont {Berg}}, \
  and\ \bibinfo {author} {\bibfnamefont {Masaki}\ \bibnamefont {Oshikawa}},\
  }\bibfield  {title} {\enquote {\bibinfo {title} {{Entanglement spectrum of a
  topological phase in one dimension}},}\ }\href {\doibase
  10.1103/PhysRevB.81.064439} {\bibfield  {journal} {\bibinfo  {journal} {Phys.
  Rev. B}\ }\textbf {\bibinfo {volume} {81}},\ \bibinfo {pages} {064439}
  (\bibinfo {year} {2010})},\ \Eprint {http://arxiv.org/abs/0910.1811}
  {arXiv:0910.1811} \BibitemShut {NoStop}%
\bibitem [{\citenamefont {Efrati}\ \emph {et~al.}(2014)\citenamefont {Efrati},
  \citenamefont {Wang}, \citenamefont {Kolan},\ and\ \citenamefont
  {Kadanoff}}]{Efrati2014}%
  \BibitemOpen
  \bibfield  {author} {\bibinfo {author} {\bibfnamefont {Efi}\ \bibnamefont
  {Efrati}}, \bibinfo {author} {\bibfnamefont {Zhe}\ \bibnamefont {Wang}},
  \bibinfo {author} {\bibfnamefont {Amy}\ \bibnamefont {Kolan}}, \ and\
  \bibinfo {author} {\bibfnamefont {Leo~P.}\ \bibnamefont {Kadanoff}},\
  }\bibfield  {title} {\enquote {\bibinfo {title} {{Real-space renormalization
  in statistical mechanics}},}\ }\href {\doibase 10.1103/RevModPhys.86.647}
  {\bibfield  {journal} {\bibinfo  {journal} {Rev. Mod. Phys.}\ }\textbf
  {\bibinfo {volume} {86}},\ \bibinfo {pages} {647--667} (\bibinfo {year}
  {2014})}\BibitemShut {NoStop}%
\bibitem [{Note3()}]{Note3}%
  \BibitemOpen
  \bibinfo {note} {One can also require the diagonal of the $R$ matrix to be
  positive to fix the redundant gauge degrees of freedom. This can be easily
  implemented on top of the existing QR libraries with backward
  support.}\BibitemShut {Stop}%
\bibitem [{\citenamefont {Chen}\ \emph {et~al.}(2016)\citenamefont {Chen},
  \citenamefont {Xu}, \citenamefont {Zhang},\ and\ \citenamefont
  {Guestrin}}]{chen2016training}%
  \BibitemOpen
  \bibfield  {author} {\bibinfo {author} {\bibfnamefont {Tianqi}\ \bibnamefont
  {Chen}}, \bibinfo {author} {\bibfnamefont {Bing}\ \bibnamefont {Xu}},
  \bibinfo {author} {\bibfnamefont {Chiyuan}\ \bibnamefont {Zhang}}, \ and\
  \bibinfo {author} {\bibfnamefont {Carlos}\ \bibnamefont {Guestrin}},\
  }\bibfield  {title} {\enquote {\bibinfo {title} {Training deep nets with
  sublinear memory cost},}\ }\href@noop {} {\bibfield  {journal} {\bibinfo
  {journal} {arXiv}\ } (\bibinfo {year} {2016})},\ \Eprint
  {http://arxiv.org/abs/1604.06174} {arXiv:1604.06174} \BibitemShut {NoStop}%
\bibitem [{\citenamefont {Christianson}(1994)}]{Christianson1994}%
  \BibitemOpen
  \bibfield  {author} {\bibinfo {author} {\bibfnamefont {Bruce}\ \bibnamefont
  {Christianson}},\ }\bibfield  {title} {\enquote {\bibinfo {title} {{Reverse
  accumulation and attractive fixed points}},}\ }\href {\doibase
  10.1080/10556789408805572} {\bibfield  {journal} {\bibinfo  {journal} {Optim.
  Methods Softw.}\ }\textbf {\bibinfo {volume} {3}},\ \bibinfo {pages}
  {311--326} (\bibinfo {year} {1994})}\BibitemShut {NoStop}%
\bibitem [{\citenamefont {Fishman}\ \emph {et~al.}(2018)\citenamefont
  {Fishman}, \citenamefont {Vanderstraeten}, \citenamefont {Zauner-Stauber},
  \citenamefont {Haegeman},\ and\ \citenamefont {Verstraete}}]{Haegeman2018a}%
  \BibitemOpen
  \bibfield  {author} {\bibinfo {author} {\bibfnamefont {M.~T.}\ \bibnamefont
  {Fishman}}, \bibinfo {author} {\bibfnamefont {L.}~\bibnamefont
  {Vanderstraeten}}, \bibinfo {author} {\bibfnamefont {V.}~\bibnamefont
  {Zauner-Stauber}}, \bibinfo {author} {\bibfnamefont {J.}~\bibnamefont
  {Haegeman}}, \ and\ \bibinfo {author} {\bibfnamefont {F.}~\bibnamefont
  {Verstraete}},\ }\bibfield  {title} {\enquote {\bibinfo {title} {{Faster
  methods for contracting infinite two-dimensional tensor networks}},}\ }\href
  {\doibase 10.1103/physrevb.98.235148} {\bibfield  {journal} {\bibinfo
  {journal} {Phys. Rev. B}\ }\textbf {\bibinfo {volume} {98}},\ \bibinfo
  {pages} {235148} (\bibinfo {year} {2018})}\BibitemShut {NoStop}%
\bibitem [{Note4()}]{Note4}%
  \BibitemOpen
  \bibinfo {note} {A related example is backward through an iterative solver
  for the dominant eigenstates $x^\ast $ of a matrix $A$. In the forward
  process, one may use the Krylov space method such as Lanczos or Arnoldi
  iterations. While for the backward pass one can leverage the fact that the
  dominant eigenvector is the fixed point of the power iteration $ f(x, A) =
  Ax/||A x||$. Therefore, one can use Eq.~(\ref {eq:fixedpoint}) to propagate
  $\protect \overline {x^\ast }$ back to $\protect \overline {A}$. Note that
  the forward and the backward iteration functions are decoupled, thus the
  backward function does not need to be the same as the forward pass, as long
  as it ensures $x^\ast $ is a fixed point.}\BibitemShut {Stop}%
\bibitem [{\citenamefont {Nocedal}\ and\ \citenamefont
  {Wright}(2006)}]{Nocedal2006}%
  \BibitemOpen
  \bibfield  {author} {\bibinfo {author} {\bibfnamefont {Jorge}\ \bibnamefont
  {Nocedal}}\ and\ \bibinfo {author} {\bibfnamefont {Stephen~J.}\ \bibnamefont
  {Wright}},\ }\href {\doibase 10.1007/978-0-387-40065-5} {\emph {\bibinfo
  {title} {{Numerical optimization}}}},\ \bibinfo {edition} {2nd}\ ed.\
  (\bibinfo  {publisher} {Springer},\ \bibinfo {year} {2006})\BibitemShut
  {NoStop}%
\bibitem [{\citenamefont {Pearlmutter}(1994)}]{pearlmutter1994fast}%
  \BibitemOpen
  \bibfield  {author} {\bibinfo {author} {\bibfnamefont {Barak~A}\ \bibnamefont
  {Pearlmutter}},\ }\bibfield  {title} {\enquote {\bibinfo {title} {Fast exact
  multiplication by the hessian},}\ }\href@noop {} {\bibfield  {journal}
  {\bibinfo  {journal} {Neural computation}\ }\textbf {\bibinfo {volume} {6}},\
  \bibinfo {pages} {147--160} (\bibinfo {year} {1994})}\BibitemShut {NoStop}%
\bibitem [{pyt()}]{pytorch}%
  \BibitemOpen
  \href@noop {} {}\bibinfo {note}
  {\href{https://github.com/pytorch/pytorch}{https://github.com/pytorch/pytorch}}\BibitemShut
  {NoStop}%
\bibitem [{aut()}]{autograd}%
  \BibitemOpen
  \href@noop {} {}\bibinfo {note}
  {\href{https://github.com/HIPS/autograd}{https://github.com/HIPS/autograd}}\BibitemShut
  {NoStop}%
\bibitem [{ten()}]{tensorflow}%
  \BibitemOpen
  \href@noop {} {}\bibinfo {note}
  {\href{https://github.com/tensorflow/tensorflow}{https://github.com/tensorflow/tensorflow}}\BibitemShut
  {NoStop}%
\bibitem [{Jax()}]{Jax}%
  \BibitemOpen
  \href@noop {} {}\bibinfo {note}
  {\href{https://github.com/google/jax}{https://github.com/google/jax}}\BibitemShut
  {NoStop}%
\bibitem [{Zyg()}]{Zygote}%
  \BibitemOpen
  \href@noop {} {}\bibinfo {note}
  {\href{https://github.com/FluxML/Zygote.jl}{https://github.com/FluxML/Zygote.jl}}\BibitemShut
  {NoStop}%
\bibitem [{\citenamefont {Onsager}(1944)}]{Onsager1944}%
  \BibitemOpen
  \bibfield  {author} {\bibinfo {author} {\bibfnamefont {Lars}\ \bibnamefont
  {Onsager}},\ }\bibfield  {title} {\enquote {\bibinfo {title} {{Crystal
  statistics. I. A two-dimensional model with an order-disorder transition}},}\
  }\href {\doibase 10.1103/PhysRev.65.117} {\bibfield  {journal} {\bibinfo
  {journal} {Phys. Rev.}\ }\textbf {\bibinfo {volume} {65}},\ \bibinfo {pages}
  {117--149} (\bibinfo {year} {1944})}\BibitemShut {NoStop}%
\bibitem [{\citenamefont {Chen}\ \emph {et~al.}(2017)\citenamefont {Chen},
  \citenamefont {Liao}, \citenamefont {Xie}, \citenamefont {Han}, \citenamefont
  {Huang}, \citenamefont {Cheng}, \citenamefont {Wei}, \citenamefont {Xie},\
  and\ \citenamefont {Xiang}}]{q-clock}%
  \BibitemOpen
  \bibfield  {author} {\bibinfo {author} {\bibfnamefont {Jing}\ \bibnamefont
  {Chen}}, \bibinfo {author} {\bibfnamefont {Hai-Jun}\ \bibnamefont {Liao}},
  \bibinfo {author} {\bibfnamefont {Hai-Dong}\ \bibnamefont {Xie}}, \bibinfo
  {author} {\bibfnamefont {Xing-Jie}\ \bibnamefont {Han}}, \bibinfo {author}
  {\bibfnamefont {Rui-Zhen}\ \bibnamefont {Huang}}, \bibinfo {author}
  {\bibfnamefont {Song}\ \bibnamefont {Cheng}}, \bibinfo {author}
  {\bibfnamefont {Zhong-Chao}\ \bibnamefont {Wei}}, \bibinfo {author}
  {\bibfnamefont {Zhi-Yuan}\ \bibnamefont {Xie}}, \ and\ \bibinfo {author}
  {\bibfnamefont {Tao}\ \bibnamefont {Xiang}},\ }\bibfield  {title} {\enquote
  {\bibinfo {title} {{Phase Transition of the q-State Clock Model: Duality and
  Tensor Renormalization}},}\ }\href {\doibase 10.1088/0256-307X/34/5/050503}
  {\bibfield  {journal} {\bibinfo  {journal} {Chinese Physics Letters}\
  }\textbf {\bibinfo {volume} {34}},\ \bibinfo {pages} {050503} (\bibinfo
  {year} {2017})}\BibitemShut {NoStop}%
\bibitem [{\citenamefont {Li}\ \emph {et~al.}(2012)\citenamefont {Li},
  \citenamefont {von Delft},\ and\ \citenamefont {Xiang}}]{Li2012}%
  \BibitemOpen
  \bibfield  {author} {\bibinfo {author} {\bibfnamefont {Wei}\ \bibnamefont
  {Li}}, \bibinfo {author} {\bibfnamefont {Jan}\ \bibnamefont {von Delft}}, \
  and\ \bibinfo {author} {\bibfnamefont {Tao}\ \bibnamefont {Xiang}},\
  }\bibfield  {title} {\enquote {\bibinfo {title} {{Efficient simulation of
  infinite tree tensor network states on the Bethe lattice}},}\ }\href
  {\doibase 10.1103/PhysRevB.86.195137} {\bibfield  {journal} {\bibinfo
  {journal} {Phys. Rev. B}\ }\textbf {\bibinfo {volume} {86}},\ \bibinfo
  {pages} {195137} (\bibinfo {year} {2012})}\BibitemShut {NoStop}%
\bibitem [{\citenamefont {Vanderstraeten}\ \emph {et~al.}(2015)\citenamefont
  {Vanderstraeten}, \citenamefont {Mari\"en}, \citenamefont {Verstraete},\ and\
  \citenamefont {Haegeman}}]{Vanderstraeten2015}%
  \BibitemOpen
  \bibfield  {author} {\bibinfo {author} {\bibfnamefont {Laurens}\ \bibnamefont
  {Vanderstraeten}}, \bibinfo {author} {\bibfnamefont {Micha\"el}\ \bibnamefont
  {Mari\"en}}, \bibinfo {author} {\bibfnamefont {Frank}\ \bibnamefont
  {Verstraete}}, \ and\ \bibinfo {author} {\bibfnamefont {Jutho}\ \bibnamefont
  {Haegeman}},\ }\bibfield  {title} {\enquote {\bibinfo {title} {{Excitations
  and the tangent space of projected entangled-pair states}},}\ }\href
  {\doibase 10.1103/PhysRevB.92.201111} {\bibfield  {journal} {\bibinfo
  {journal} {Phys. Rev. B}\ }\textbf {\bibinfo {volume} {92}},\ \bibinfo
  {pages} {201111} (\bibinfo {year} {2015})}\BibitemShut {NoStop}%
\bibitem [{\citenamefont {Xie}\ \emph {et~al.}(2017)\citenamefont {Xie},
  \citenamefont {Liao}, \citenamefont {Huang}, \citenamefont {Xie},
  \citenamefont {Chen}, \citenamefont {Liu},\ and\ \citenamefont
  {Xiang}}]{Xie2017}%
  \BibitemOpen
  \bibfield  {author} {\bibinfo {author} {\bibfnamefont {Z.~Y.}\ \bibnamefont
  {Xie}}, \bibinfo {author} {\bibfnamefont {H.~J.}\ \bibnamefont {Liao}},
  \bibinfo {author} {\bibfnamefont {R.~Z.}\ \bibnamefont {Huang}}, \bibinfo
  {author} {\bibfnamefont {H.~D.}\ \bibnamefont {Xie}}, \bibinfo {author}
  {\bibfnamefont {J.}~\bibnamefont {Chen}}, \bibinfo {author} {\bibfnamefont
  {Z.~Y.}\ \bibnamefont {Liu}}, \ and\ \bibinfo {author} {\bibfnamefont
  {T.}~\bibnamefont {Xiang}},\ }\bibfield  {title} {\enquote {\bibinfo {title}
  {{Optimized contraction scheme for tensor-network states}},}\ }\href
  {\doibase 10.1103/PhysRevB.96.045128} {\bibfield  {journal} {\bibinfo
  {journal} {Phys. Rev. B}\ }\textbf {\bibinfo {volume} {96}},\ \bibinfo
  {pages} {045128} (\bibinfo {year} {2017})}\BibitemShut {NoStop}%
\bibitem [{\citenamefont {Sandvik}(2010)}]{Sandvik2010a}%
  \BibitemOpen
  \bibfield  {author} {\bibinfo {author} {\bibfnamefont {Anders~W}\
  \bibnamefont {Sandvik}},\ }\bibfield  {title} {\enquote {\bibinfo {title}
  {{Computational studies of quantum spin systems}},}\ }in\ \href {\doibase
  10.1063/1.3518900} {\emph {\bibinfo {booktitle} {AIP Conf. Proc.}}},\ Vol.\
  \bibinfo {volume} {1297}\ (\bibinfo {year} {2010})\ pp.\ \bibinfo {pages}
  {135--338}\BibitemShut {NoStop}%
\bibitem [{Note5()}]{Note5}%
  \BibitemOpen
  \bibinfo {note} {Although it was expected that the full update method will in
  principle reach the same accuracy as the variational method within the error
  of Trotter-Suzuki decomposition, it would require optimizing the iPEPS
  tensors in a globally optimal way.}\BibitemShut {Stop}%
\bibitem [{\citenamefont {Corboz}\ \emph {et~al.}(2011)\citenamefont {Corboz},
  \citenamefont {White}, \citenamefont {Vidal},\ and\ \citenamefont
  {Troyer}}]{Corboz2011}%
  \BibitemOpen
  \bibfield  {author} {\bibinfo {author} {\bibfnamefont {Philippe}\
  \bibnamefont {Corboz}}, \bibinfo {author} {\bibfnamefont {Steven~R.}\
  \bibnamefont {White}}, \bibinfo {author} {\bibfnamefont {Guifr{\'{e}}}\
  \bibnamefont {Vidal}}, \ and\ \bibinfo {author} {\bibfnamefont {Matthias}\
  \bibnamefont {Troyer}},\ }\bibfield  {title} {\enquote {\bibinfo {title}
  {{Stripes in the two-dimensional t-J model with infinite projected
  entangled-pair states}},}\ }\href {\doibase 10.1103/PhysRevB.84.041108}
  {\bibfield  {journal} {\bibinfo  {journal} {Phys. Rev. B}\ }\textbf {\bibinfo
  {volume} {84}},\ \bibinfo {pages} {041108} (\bibinfo {year}
  {2011})}\BibitemShut {NoStop}%
\bibitem [{\citenamefont {Liao}\ \emph {et~al.}(2017)\citenamefont {Liao},
  \citenamefont {Xie}, \citenamefont {Chen}, \citenamefont {Liu}, \citenamefont
  {Xie}, \citenamefont {Huang}, \citenamefont {Normand},\ and\ \citenamefont
  {Xiang}}]{Liao2017}%
  \BibitemOpen
  \bibfield  {author} {\bibinfo {author} {\bibfnamefont {H.~J.}\ \bibnamefont
  {Liao}}, \bibinfo {author} {\bibfnamefont {Z.~Y.}\ \bibnamefont {Xie}},
  \bibinfo {author} {\bibfnamefont {J.}~\bibnamefont {Chen}}, \bibinfo {author}
  {\bibfnamefont {Z.~Y.}\ \bibnamefont {Liu}}, \bibinfo {author} {\bibfnamefont
  {H.~D.}\ \bibnamefont {Xie}}, \bibinfo {author} {\bibfnamefont {R.~Z.}\
  \bibnamefont {Huang}}, \bibinfo {author} {\bibfnamefont {B.}~\bibnamefont
  {Normand}}, \ and\ \bibinfo {author} {\bibfnamefont {T.}~\bibnamefont
  {Xiang}},\ }\bibfield  {title} {\enquote {\bibinfo {title} {{Gapless
  Spin-Liquid Ground State in the S=1 /2 Kagome Antiferromagnet}},}\ }\href
  {\doibase 10.1103/PhysRevLett.118.137202} {\bibfield  {journal} {\bibinfo
  {journal} {Phys. Rev. Lett.}\ }\textbf {\bibinfo {volume} {118}},\ \bibinfo
  {pages} {137202} (\bibinfo {year} {2017})}\BibitemShut {NoStop}%
\bibitem [{\citenamefont {Lee}\ \emph {et~al.}(2018)\citenamefont {Lee},
  \citenamefont {Normand},\ and\ \citenamefont {Kao}}]{Lee2018f}%
  \BibitemOpen
  \bibfield  {author} {\bibinfo {author} {\bibfnamefont {Chih-Yuan}\
  \bibnamefont {Lee}}, \bibinfo {author} {\bibfnamefont {B.}~\bibnamefont
  {Normand}}, \ and\ \bibinfo {author} {\bibfnamefont {Ying-Jer}\ \bibnamefont
  {Kao}},\ }\bibfield  {title} {\enquote {\bibinfo {title} {{Gapless spin
  liquid in the kagome Heisenberg antiferromagnet with Dzyaloshinskii-Moriya
  interactions}},}\ }\href {\doibase 10.1103/PhysRevB.98.224414} {\bibfield
  {journal} {\bibinfo  {journal} {Phys. Rev. B}\ }\textbf {\bibinfo {volume}
  {98}},\ \bibinfo {pages} {224414} (\bibinfo {year} {2018})},\ \Eprint
  {http://arxiv.org/abs/1809.09128} {arXiv:1809.09128} \BibitemShut {NoStop}%
\bibitem [{\citenamefont {Haghshenas}\ \emph {et~al.}(2018)\citenamefont
  {Haghshenas}, \citenamefont {Gong},\ and\ \citenamefont
  {Sheng}}]{Haghshenas2018}%
  \BibitemOpen
  \bibfield  {author} {\bibinfo {author} {\bibfnamefont {R.}~\bibnamefont
  {Haghshenas}}, \bibinfo {author} {\bibfnamefont {Shou-Shu}\ \bibnamefont
  {Gong}}, \ and\ \bibinfo {author} {\bibfnamefont {D.~N.}\ \bibnamefont
  {Sheng}},\ }\bibfield  {title} {\enquote {\bibinfo {title} {{An iPEPS study
  of kagome Heisenberg model with chiral interaction: A single-layer
  tensor-network algorithm}},}\ }\href {http://arxiv.org/abs/1812.11436}
  {\bibfield  {journal} {\bibinfo  {journal} {arXiv}\ } (\bibinfo {year}
  {2018})},\ \Eprint {http://arxiv.org/abs/1812.11436} {arXiv:1812.11436}
  \BibitemShut {NoStop}%
\bibitem [{\citenamefont {Vanderstraeten}\ \emph {et~al.}(2018)\citenamefont
  {Vanderstraeten}, \citenamefont {Vanhecke},\ and\ \citenamefont
  {Verstraete}}]{Vanderstraeten2018a}%
  \BibitemOpen
  \bibfield  {author} {\bibinfo {author} {\bibfnamefont {Laurens}\ \bibnamefont
  {Vanderstraeten}}, \bibinfo {author} {\bibfnamefont {Bram}\ \bibnamefont
  {Vanhecke}}, \ and\ \bibinfo {author} {\bibfnamefont {Frank}\ \bibnamefont
  {Verstraete}},\ }\bibfield  {title} {\enquote {\bibinfo {title} {{Residual
  entropies for three-dimensional frustrated spin systems with tensor
  networks}},}\ }\href {\doibase 10.1103/PhysRevE.98.042145} {\bibfield
  {journal} {\bibinfo  {journal} {Phys. Rev. E}\ }\textbf {\bibinfo {volume}
  {98}},\ \bibinfo {pages} {042145} (\bibinfo {year} {2018})},\ \Eprint
  {http://arxiv.org/abs/1805.10598} {arXiv:1805.10598} \BibitemShut {NoStop}%
\bibitem [{\citenamefont {Ionescu}\ \emph {et~al.}(2015)\citenamefont
  {Ionescu}, \citenamefont {Vantzos},\ and\ \citenamefont
  {Sminchisescu}}]{Ionescu2015}%
  \BibitemOpen
  \bibfield  {author} {\bibinfo {author} {\bibfnamefont {Catalin}\ \bibnamefont
  {Ionescu}}, \bibinfo {author} {\bibfnamefont {Orestis}\ \bibnamefont
  {Vantzos}}, \ and\ \bibinfo {author} {\bibfnamefont {Cristian}\ \bibnamefont
  {Sminchisescu}},\ }\bibfield  {title} {\enquote {\bibinfo {title} {{Training
  Deep Networks with Structured Layers by Matrix Backpropagation}},}\ }in\
  \href {http://arxiv.org/abs/1509.07838} {\emph {\bibinfo {booktitle} {Proc.
  IEEE Int. Conf. Comput. Vis.}}}\ (\bibinfo {year} {2015})\ pp.\ \bibinfo
  {pages} {2965--2973},\ \Eprint {http://arxiv.org/abs/1509.07838}
  {arXiv:1509.07838} \BibitemShut {NoStop}%
\bibitem [{\citenamefont {Novikov}\ \emph {et~al.}(2015)\citenamefont
  {Novikov}, \citenamefont {Podoprikhin}, \citenamefont {Osokin},\ and\
  \citenamefont {Vetrov}}]{novikov2015tensorizing}%
  \BibitemOpen
  \bibfield  {author} {\bibinfo {author} {\bibfnamefont {Alexander}\
  \bibnamefont {Novikov}}, \bibinfo {author} {\bibfnamefont {Dmitrii}\
  \bibnamefont {Podoprikhin}}, \bibinfo {author} {\bibfnamefont {Anton}\
  \bibnamefont {Osokin}}, \ and\ \bibinfo {author} {\bibfnamefont {Dmitry~P}\
  \bibnamefont {Vetrov}},\ }\bibfield  {title} {\enquote {\bibinfo {title}
  {Tensorizing neural networks},}\ }in\ \href@noop {} {\emph {\bibinfo
  {booktitle} {Advances in neural information processing systems}}}\ (\bibinfo
  {year} {2015})\ pp.\ \bibinfo {pages} {442--450}\BibitemShut {NoStop}%
\bibitem [{\citenamefont {Laue}\ \emph {et~al.}(2018)\citenamefont {Laue},
  \citenamefont {Mitterreiter},\ and\ \citenamefont {Giesen}}]{NIPS2018_7540}%
  \BibitemOpen
  \bibfield  {author} {\bibinfo {author} {\bibfnamefont {Soeren}\ \bibnamefont
  {Laue}}, \bibinfo {author} {\bibfnamefont {Matthias}\ \bibnamefont
  {Mitterreiter}}, \ and\ \bibinfo {author} {\bibfnamefont {Joachim}\
  \bibnamefont {Giesen}},\ }\bibfield  {title} {\enquote {\bibinfo {title}
  {Computing higher order derivatives of matrix and tensor expressions},}\ }in\
  \href
  {http://papers.nips.cc/paper/7540-computing-higher-order-derivatives-of-matrix-and-tensor-expressions.pdf}
  {\emph {\bibinfo {booktitle} {Advances in Neural Information Processing
  Systems 31}}}\ (\bibinfo {year} {2018})\ pp.\ \bibinfo {pages}
  {2750--2759}\BibitemShut {NoStop}%
\end{thebibliography}%

\clearpage
\appendix
\end{document}